\documentclass[12pt, letterpaper]{article}

\pdfoutput=1

\usepackage{amsmath}
\usepackage{amsfonts}
\usepackage{amssymb}
\usepackage[colorlinks,citecolor=blue,urlcolor=blue,bookmarks=false,hypertexnames=true]{hyperref}
\usepackage{cite}

\usepackage{graphicx, physics, subcaption}
\usepackage{verbatim }

\numberwithin{equation}{section}

\usepackage{color}

\usepackage{amsmath, amssymb, graphics, epsfig, graphicx}
\usepackage{epsf}
\usepackage{epstopdf}
\usepackage {amssymb}
\newcommand{\nc}{\newcommand}

\newcommand{\calR}{{\cal{R}}}
\newcommand{\calP}{{\cal{P}}}

\nc{\cN}{ {\cal{N}} }

\nc{\ba}{\begin{eqnarray}}
\nc{\ea}{\end{eqnarray}}

\begin{document}

\vspace{5mm}
\vspace{0.5cm}
\begin{center}

\def\thefootnote{\fnsymbol{footnote}}

{\bf\large Stochastic Multiple Fields Inflation: Diffusion Dominated Regime}
\\[0.5cm]

{Kosar Asadi \footnote{k.asadi@ipm.ir}, 
Amin Nassiri-Rad \footnote{amin.nassiriraad@ipm.ir}, 
Hassan Firouzjahi \footnote{firouz@ipm.ir},
}
\\[.7cm]

{\small \textit{$^{1}$School of Astronomy, 
		Institute for Research in Fundamental Sciences (IPM) \\ 
		P.~O.~Box 19395-5531, Tehran, Iran }} \\

\end{center}

\vspace{.8cm}

\hrule \vspace{0.3cm}


\begin{abstract}

We study multiple fields inflation in diffusion dominated regime using stochastic $\delta N$ formalism. The fields are under pure Brownian motion in a 
dS background with boundaries in higher dimensional  field space.  
This setup can be realized towards the final stages of the ultra slow-roll setup where the classical drifts  fall off exponentially and the perturbations are driven by
quantum kicks. We consider both symmetric and asymmetric boundaries with absorbing and reflective boundary conditions and calculate the average number of e-folds, the first crossing probabilities and the power spectrum. We study the primordial black holes (PBHs) formation in this setup and calculate the mass fraction and the contribution of PBHs in dark matter energy density for various higher dimensional field spaces.

\end{abstract}
\vspace{0.5cm} \hrule
\def\thefootnote{\arabic{footnote}}
\setcounter{footnote}{0}
\newpage

\section{Introduction}

The simplest models of cosmic inflation are driven by a single scalar field, the inflaton field,   which slowly rolls on top of its nearly flat potential \cite{Weinberg:2008zzc, Baumann:2009ds}. To solve the flatness and the horizon problems, one typically requires that inflation lasts for nearly 60 e-folds or so. Among the key predictions of these simple models of inflation are that the primordial perturbations generated from the quantum fluctuations of the inflaton field are nearly scale invariant, adiabatic and Gaussian, which are well consistent with cosmological observations \cite{Planck:2018vyg, Planck:2018jri}. While single field scenarios are most economical from the model building point of view, but multiple fields models of inflation are natural as well. Indeed,  theories of high energy physics typically contain many fields in their spectrum. If some of these fields are light enough in early Universe  they can drive inflation collectively yielding to multiple field scenarios of inflation, see  \cite{Wands:2007bd, Dimopoulos:2005ac, Ashoorioon:2009wa} for examples.

Cosmological perturbation theory is the standard method to study perturbations generated during inflation \cite{Kodama:1984ziu}. One can use the perturbative QFT methods such as the in-in formalism to calculate various correlations such as the power spectrum and bispectrum of curvature perturbations \cite{Weinberg:2005vy}. Alternatively, one can use the method of stochastic inflation to study cosmological perturbations during inflation \cite{Vilenkin:1983xp, Starobinsky:1986fx, Starobinsky:1994bd}. 
Stochastic inflation is an effective approach to study the dynamics of the long superhorizon perturbations which are  affected by small scale quantum perturbations.  More specifically,  one decomposes the cosmological perturbations into the long and short perturbations. As the Universe expands the 
short modes are stretched beyond the horizon and  become classical which  affect the dynamics of long mode perturbations. The effects of short modes on long modes are described by random classical white noises with the amplitude $\frac{H}{2 \pi}$ in which $H$ is the Hubble expansion rate during inflation. The formalism of stochastic inflation has been used  in slow-roll models \cite{ Rey:1986zk, Nakao:1988yi, Sasaki:1987gy, Nambu:1987ef, Nambu:1988je,  Kandrup:1988sc,  Nambu:1989uf, Mollerach:1990zf, Linde:1993xx, Starobinsky:1994bd, Kunze:2006tu, Prokopec:2007ak, Prokopec:2008gw, Tsamis:2005hd, Enqvist:2008kt, 
Finelli:2008zg, Finelli:2010sh, Garbrecht:2013coa, Garbrecht:2014dca, Burgess:2014eoa,  Burgess:2015ajz, Boyanovsky:2015tba,  Boyanovsky:2015jen, Pinol:2020cdp, Cruces:2018cvq, Cruces:2021iwq, Noorbala:2019kdd, Ahmadi:2022lsm},  ultra slow-roll setups \cite{Pattison:2019hef, Pattison:2021oen, Firouzjahi:2018vet, Firouzjahi:2020jrj, Mishra:2023lhe} and also in models involving gauge fields \cite{ Fujita:2017lfu, Fujita:2022fit, Fujita:2022fwc, Talebian:2019opf, Talebian:2020drj, Talebian:2021dfq, Talebian:2022jkb}.  To calculate curvature perturbation and its correlations one implement $\delta N$ method in stochastic formalism \cite{Fujita:2013cna, Fujita:2014tja, Vennin:2015hra, Vennin:2016wnk, Assadullahi:2016gkk,  Grain:2017dqa, Noorbala:2018zlv, Jackson:2022unc}.  The $\delta N$ formalism is based on the separate universe approach where the superhorizon perturbations modify  the background expansion of the nearby patches \cite{Sasaki:1995aw, Sasaki:1998ug, Lyth:2004gb, Wands:2000dp, Lyth:2005fi, Abolhasani:2019cqw, Abolhasani:2018gyz}. Since in stochastic formalism one decomposes the perturbations into  long and short modes this allows one to  employ $\delta N$ formalism to study  the superhorizon perturbations.

Quantum  fluctuations generated during inflation 
are usually described by curvature perturbations $\calR$. In the range of scales accessible to CMB observations,  these fluctuations are constrained to be at the order $\calR \sim 10^{-5}$.  On the other hand, at smaller scales $\calR$ can grow 
by few orders of magnitude, say $\calR \sim 10^{-1}$ to seed the primordial black holes (PBHs) formation \cite{Hawking:1971mnras, Carr:1974 mnras, CarrPBH:1975 Apj, Sasaki:2018dmp, Carr:2020gox, Green:2020jor, Byrnes:2021jka} as a candidate for dark matter or gravitational waves events detected by LIGO/VIRGO observations. Models of ultra slow-roll (USR) inflation \cite{Kinney:2005vj} are well studied in recent years as a setup to generate PBHs. This is because in USR setups the potential is very flat so curvature perturbations can grow on superhorizon scales \cite{Namjoo:2012aa, Chen:2013aj, Martin:2012pe, Akhshik:2015rwa}. As the potential is flat, one expects that the quantum diffusion effects to play important roles during USR setup. This was studied in single field USR setup in  \cite{Pattison:2019hef, Pattison:2021oen, Firouzjahi:2018vet, Firouzjahi:2020jrj}. Motivated by these studies, in this work we study stochastic effects in models of multiple fields inflation where the system is diffusion dominated. The limit of diffusion domination takes place  at the final stage of USR setup where the fields' velocities are exponentially damped and the corresponding Langevin equations   are dominated by the quantum diffusion terms. This work is an  extension of the stochastic multiple fields inflation setup studied in \cite{ Vennin:2016wnk, Assadullahi:2016gkk} in the drift dominated regime.  For a recent work concerning single field diffusion dominated setup with stochastic boundaries see \cite{Nassiri-Rad:2022azj}.

As the system is diffusion dominated, the dynamics of the quantum fluctuations of the fields are given by pure Brownian motion in higher dimensional field space.  The random Brownian motions are subject to  boundaries in field space, both in UV and IR regions. The former boundary is a realization of the fact that the field may not probe arbitrary high energy region of the field space while the latter boundary may  represent the surface of end of inflation.  
With a given initial condition in field space, the number of e-folds $\cN$ when the fields hit either of the boundaries is a stochastic variable which plays important roles in our analysis below.

The rest of the  paper is organized as follows. In section \ref{symmetric} we study  a two-dimensional model with concentric circular boundaries  and then extend these analyses to $n>2$ dimensional field space  with concentric $n-1$ dimensional spherical boundaries in  field space. In Section \ref{asymmetry} we repeat these analyses to the setup with asymmetric boundaries in two dimensions. In Section \ref{PBH} we look at the predictions of the model for PBHs formation followed by summary and discussions in Section \ref{conclusion}. Some technicalities associated to Fokker-Planck equation is relegated to the Appendix. 

\section{Symmetric Boundaries}
\label{symmetric}

As described above, we study the quantum fluctuations of multiple fields in diffusion dominated regime when the dynamics of the fields are governed by pure Brownian motion in field space. 
There are a number of motivations to study this regime. The assumption of diffusion domination regime is well justified towards the end of USR phase where the classical velocities of the fields fall off exponentially and one can neglect the classical drift terms in the Langevin equations. 
For example, consider the Langevin equation in the phase space for the light single field  $\phi$ in the USR setup  \cite{Firouzjahi:2018vet}
\ba
 \frac{d\phi}{d N}= \frac{v}{H} +\frac{H}{2 \pi}\xi(N) ,\,\,\,\,\,\,\,\, \frac{d v}{d N}=-3 v \, ,
\ea  
in which $N$ is the number of e-folds which is used as a clock, $H$ is the Hubble expansion rate during inflation  and $v(N)$ is the usual conjugate momentum associated to the field $\phi$. In addition, $\xi$ is the Gaussian white noise with the unit amplitude. 
As shown in \cite{Firouzjahi:2018vet}, the conjugate momentum $v(N)$ falls off exponentially, $v(N) = v_0 e^{-3N}$ so if one waits long enough, the effects of the classical drift in the Langevin equation of $\phi$ is neglected and the system is given by a pure Brownian motion 
\ba
\label{brownian}
 \frac{d\phi}{d N}=\frac{H}{2 \pi}\xi(N) \, .
\ea 
In this work we extend Eq. (\ref{brownian}) to multiple fields setup with boundaries in field space.   This corresponds to the case when multiple light fields drive the USR phase of inflation.  To simplify the analysis below, we rescale all fields in the unit of $\frac{H}{2 \pi}$ so all fields are dimensionless. 

In order to terminate/restrict the pure Brownian motion  we impose boundaries in field space. The boundaries can be either reflective or absorbing. In the former case, the boundary forbids the field to go further and reflect it back so the field 
resumes anew its dynamics after hitting the boundary while in the latter the dynamics of the field is terminated. 
The absorbing boundaries may represent the surface of end of USR (or end of inflation) while the reflective boundaries may represent the forbidden regime of field space where the fields can not probe. For example, the very large field limit, $\phi \gg M_P$ may be disallowed by our semi-quantum treatment of inflation so  reflective boundaries may be inserted in the UV region of field space 
to forbid the quantum gravity limit.    

In this Section we study the cases of symmetric boundaries in two and higher dimensional field space where analytic results can be obtained. On the other hands, for asymmetric boundaries, it is not easy to find compact analytic results. So 
in Section \ref{asymmetry} we study various asymmetric boundaries in two dimensions.


\subsection{Two Dimensional Circular Boundary}
\label{Circle}

\begin{figure}
 \vspace{-0.75cm}
    \centering
   \includegraphics[scale=0.5]{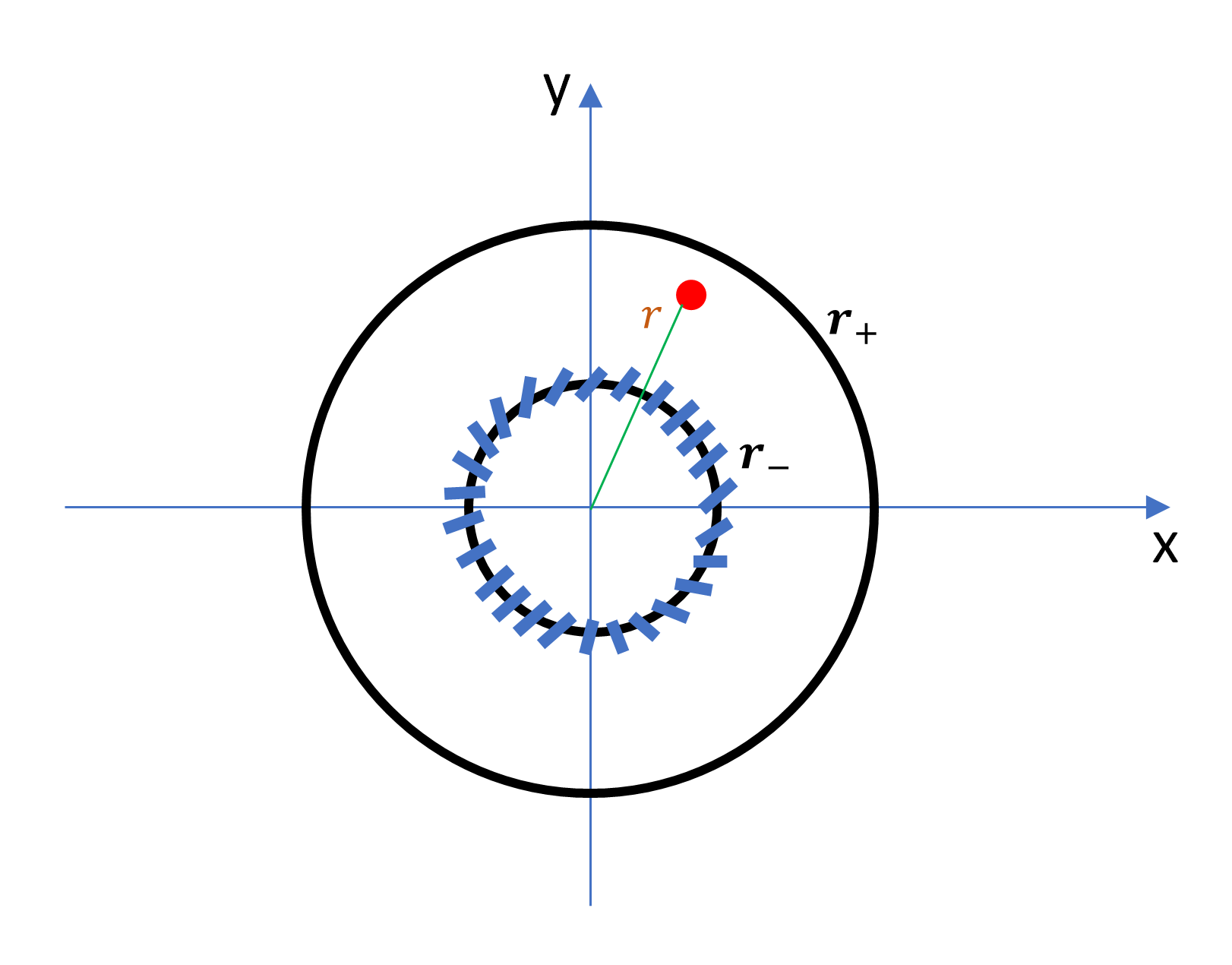}
    \caption{A schematic view of the two-dimensional circular boundaries. In this figure the inner boundary is reflective while the outer one is absorbing.}
    \label{Boundariescir}
\end{figure}

In this section we consider a pure Brownian motion in two dimensional field space. 
As we will see from our general analysis for an $n$-dimensional field space, the 
case $n=2$ is somewhat the special limit which deserves a separate investigation. 
The boundaries are two concentric circles which can be either absorbing or reflective. Fig. \ref{Boundariescir} illustrates a schematic view of this setup 
 in which the inner boundary is reflective while the outer one is absorbing.

The corresponding Langevin equation describing the Brownian motion of two random fields (with the fields and the noises normalized in the unit of $\frac{H}{2 \pi}$)  are given by 
\ba
\label{twodimensionalLangevin}
    \frac{d\phi_1}{d N}=\xi_1(N) \, , \quad \quad 
    \frac{d\phi_2}{d N}= \xi_2 (N),
    \ea
in which $\xi_i(N), i=1, 2$ are two Gaussian random noises satisfying
\ba
\langle \xi_i(N) \xi_j(N') \rangle = \delta_{ij} \delta (N- N') \, .
\ea

A key role is played by the stochastic variable $\cN$ describing the time (i.e. number of e-folds) when the fields hit either of boundaries.  We can calculate $\langle \cN \rangle$ and $\delta \cN^2 \equiv \langle  \cN^2 \rangle- \langle  \cN \rangle^2$ to obtain the power spectrum of curvature perturbations using the stochastic $\delta N$ formalism. The probability density function (PDF) $P(\mathcal{N},\phi_1,\phi_2)$
associated to the above Langevin equations is governed by the following 
adjoint Fokker-Planck equation \cite{ Assadullahi:2016gkk}
\begin{equation}
 \frac{\partial P(\mathcal{N},\phi_1,\phi_2)}{\partial \mathcal{N}}
 =\frac{1}{2} \nabla^2 P(\mathcal{N},\phi_1,\phi_2) \, ,
\end{equation}
in which $\nabla^2$ is the Laplacian in two dimensional field space. 

We impose the symmetric circular boundaries in field space given by 
\begin{equation}
    \phi_1^2+\phi_2^2=r_\pm^2\,,
\end{equation}
where without loss of generality we have assumed that the center of circles is set at the origin.  The initial condition of the fields can be taken as 
$(\phi_1,\phi_2)$ at a distance of $r=\sqrt{\phi_{1}^2+\phi_{2}^2}$ where $r$ is somewhere between $r_+$ and $r_-$.
As the boundaries have the circular symmetry it is more convenient to use the polar coordinate $(r, \theta)$ and write the Laplacian in the adjoint Fokker-Planck equation as
\begin{equation}\label{adjointfokker}
    \frac{1}{2 r}\frac{\partial}{\partial r}\Big(r\frac{\partial P(\mathcal{N},r,\theta )}{\partial r} \Big)+\frac{1}{2r^2}\frac{\partial^2 P(\mathcal{N},r,\theta)}{\partial\theta^2}=\frac{\partial P(\mathcal{N},r, \theta)}{\partial \mathcal{N}} \, .
\end{equation}

In order to terminate inflation we need at least one of the boundaries to be absorbing. 
If we assume  both of the boundaries to be absorbing then the corresponding boundary conditions of $P(\mathcal{N},r,\theta)$ are given by \cite{Pattison:2021oen}
\begin{equation}\label{absorbingcond}
P(\mathcal{N},r_-,\theta)=P(\mathcal{N},r_+,\theta)=\delta(\mathcal{N}) \, .
\end{equation}
On the other hand, if we assume the inner boundary $r= r_-$ to be absorbing while the outer boundary $r=r_+$ is reflecting, then the corresponding boundary conditions are 
\begin{equation}\label{reflectivecond}
   \frac{\partial P(\mathcal{N},r,\theta)}{\partial r}\Big|_{r=r_+}=0
\end{equation}
\begin{equation}
P(\mathcal{N},r_-,\theta)=\delta(\mathcal{N}) \, .
\end{equation}
If one switches the positions of the reflective and the absorbing boundaries from 
$r_{\pm}$ to $r_{\mp}$ then the corresponding boundary conditions switch as well. Correspondingly, as we shall see below, the results in this case are easily obtained by switching $r_{\pm}$ to $r_{\mp}$ in the corresponding expressions.

Since solving the  adjoint Fokker-Planck equation with the above boundary conditions can be hard, it is more convenient to use the Fourier transformation of $P(\mathcal{N},r,\theta)$ known as the characteristic function which is defined as \cite{Pattison:2021oen}
\begin{equation}\label{Char1}
    \chi_\mathcal{N}(t,r,\theta)=\int_{-\infty}^\infty e^{i t \mathcal{N}}P(\mathcal{N},r,\theta)d\mathcal{N} \, .
\end{equation}
Taking the Fourier transformation of Eq. \eqref{adjointfokker}, we rewrite it in terms of characteristic function as follows
\begin{equation}\label{Fourieradjioint}
    \frac{1}{2 r}\frac{\partial}{\partial r}\Big(r\frac{\partial \chi_{\mathcal{N}}(t,r,\theta)}{\partial r} \Big)+\frac{1}{2r^2}\frac{\partial^2 \chi_{\mathcal{N}}(t,r,\theta)}{\partial\theta^2}=-it\chi_{\mathcal{N}}(t,r,\theta) \, .
\end{equation}
The absorbing boundary conditions \eqref{absorbingcond} then become
\begin{equation}\label{charchteristicboundary}
    \chi_\mathcal{N}(t,r_-,\theta)=\chi_\mathcal{N}(t,r_+,\theta)=1 \, ,
\end{equation}
while the reflective boundary condition \eqref{reflectivecond} is written as
\begin{equation}\label{charchteristicboundary2}
     \frac{\partial \chi_\mathcal{N}(t,r,\theta)}{\partial r}\Big|_{r=r_+}=0 \, .
\end{equation}
Solving the eigenvalue equation of \eqref{Fourieradjioint} one can  calculate the moments of $\mathcal{N}$ such as $\langle \cN \rangle, \langle \cN^2 \rangle$ etc. 

From the homogeneities of the boundary conditions and the fact that the boundaries are spherical symmetric one can easily show that the characteristic function does not depend on polar angle $\theta$ and the general solution of Eq. \eqref{Fourieradjioint} is given by
\begin{equation}\label{charcircle}
    \chi_\mathcal{N}(t,r,\theta)=A J_0(\sqrt{2 i t}r)+B Y_0(\sqrt{2 i t}r),
\end{equation}
where $J_0(x)$ and $Y_0(x)$ are the first and second kind Bessel functions respectively while $A$ and $B$ are two constants determined from the boundary conditions.

\begin{itemize}
  \item \textbf{Absorbing boundaries}
\end{itemize}
First consider both boundaries to be absorbing. Using Eq. \eqref{charchteristicboundary} one can easily show that $A$ and $B$ are given by 
\begin{equation}\label{Aabsorb}
   A= -\frac{Y_0\left( \sqrt{2i t} r_+\right)-Y_0\left( \sqrt{2 i t} r_-\right)}{J_0\left( \sqrt{2i t} r_+\right) Y_0\left( \sqrt{2 i t} r_-\right)-J_0\left( \sqrt{2 i t} r_-\right) Y_0\left( \sqrt{2 i t} r_+\right)} \, ,
\end{equation}
and
\begin{equation}
  B=-\frac{J_0\left(\ \sqrt{2i t} r_+\right)-J_0\left( \sqrt{2i t} r_-\right)}{J_0\left( \sqrt{2i t} r_-\right) Y_0\left( \sqrt{2i t} r_+\right)-J_0\left( \sqrt{2i t} r_+\right) Y_0\left( \sqrt{2i t} r_-\right)} \, .
\end{equation}
Using the above equations one can show that $\chi_\mathcal{N}(t=0,r,\phi)=1$ which guarantees that $P(\mathcal{N},r,\phi)$ is normalized. 

Having the characteristic function at hand, the moments $\langle \cN^s \rangle$ with $s=1, 2,...$
are calculated as follows 
\cite{Pattison:2021oen}
\begin{equation}\label{timetwoboundary}
    \big \langle \cN ^s \big\rangle= i^{-s} \frac{\partial^s }{\partial t^s} \chi_\mathcal{N}(t,r,\phi)\big|_{t=0} \, .
\end{equation}
In particular, the average time needed to hit the boundaries with $s=1$ in 
Eq. (\ref{timetwoboundary}) is calculated as  
\begin{equation}
\label{timetwoboundary2}
    \big < \mathcal{N}\big> = \frac{r_-^2 \ln \left(\frac{r_+}{r}\right)+r_+^2 \ln \left(\frac{r}{r_-}\right)}{2 \ln \left(\frac{r_+}{r_-}\right)}  -\frac{r^2}{2}\, ,
\end{equation}
in which $r$ is the initial position of the field. 
As one expects, the above expression goes to zero if we set $r$ equal to either of the boundaries. It is also interesting to note that the above expression reduces to the following form if we set $r_-=0$,
\begin{equation}\label{toneboundary}
  \big< \mathcal{N}\big>=\frac{1}{2} \left(r_+^2-r^2\right) \, , \quad \quad (r_-=0) \, .
\end{equation}
It is worth noting that if  we set $r=0$ the above expression yields  $\big< \mathcal{N}\big>= r_+^2/2$ which is the hallmark of the pure Brownian motion in which the average time required to traverse a given length in random walks is  proportional to the square of the length. The proportionality constant depends on the dimensionality of field space, which here  is 2. 

Moreover one can calculate  the first hitting probabilities, i.e. the probability $p_+ (p_-)$ of hitting $r_+$ ($r_-$) earlier than $r_-$ ($r_+$) to terminate inflation.  To this end we start with the integral solution of Eq. \eqref{twodimensionalLangevin}
\begin{equation}
\label{sol-formal}
    \tilde{\phi}_i=\phi_i+W_i(\cN) \, , \quad \quad (i=1, 2) , 
\end{equation}
where $W_i(N)$ is the Wiener process defined by
\ba
W_i(\cN) \equiv \int_0^{\cN} \xi_i(N) d N \, ,
\ea 
satisfying the following relations,
\ba
\label{Weiner-rel}
\big<W(\mathcal{N})\big>=0  , \quad  \big<W(\mathcal{N})^2\big>=\big<\mathcal{N}\big> \, .
\ea

Now by squaring both sides of Eq. (\ref{sol-formal}) and using the relations (\ref{Weiner-rel}) one obtains,
\begin{equation}
    \big< \tilde{\phi}_1^2+\tilde{\phi}_2^2\big>=r^2+2\big<\mathcal{N}\big> \, .
\end{equation}
 On the other hand, from the definition of first hitting probabilities $p_\pm$ we have 
 $   \big< \tilde{\phi}_1^2+\tilde{\phi}_2^2\big>=p_+ r_+^2+p_- r_-^2$ as well as $p_++p_-=1$. Using these relations along with  Eq. (\ref{timetwoboundary2}) for $\langle \cN \rangle$,  one can easily obtain $p_\pm$ as follows, 
\begin{equation}
\label{ppm2}
    p_+=\frac{\ln(\frac{r}{r_-})}{\ln(\frac{r_+}{r_-})} ,\quad \quad 
    p_-=\frac{\ln \Big(\frac{r_+}{r} \Big)}{\ln(\frac{r_+}{r_-})} \, .
\end{equation}
It is worth mentioning that $p_\pm$ depends logarithmically on the position of the other boundary $r_\mp$. Furthermore, if we set $r_-=0$ then $p_+=1$ which we expect since the only outcome for the field is to hit the outer boundary.  In this limit from equation $p_+ r_+^2+p_- r_-^2=r^2+2\big<\mathcal{N}\big>$ one obtains  \eqref{toneboundary}  for $\big<\mathcal{N}\big>$ as expected.

\begin{itemize}
  \item \textbf{Mixed boundaries}
\end{itemize}

Now suppose that one of the boundaries, i.e $r_+$ is reflecting and $r_-$ is absorbing. In this case using Eq. \eqref{charchteristicboundary} for $r_-$ and Eq. \eqref{charchteristicboundary2} for $r_+$ the constants $A$ and $B$ are obtained as
\begin{equation}\label{coefref-A-circle}
   A= \frac{Y_1\left( \sqrt{2 i t} r_+\right)}{J_0\left( \sqrt{2 i t} r_-\right) Y_1\left( \sqrt{2 i t} r_+\right)-J_1\left( \sqrt{2i t} r_+\right) Y_0\left( \sqrt{2i t} r_-\right)}
\end{equation}
and
\begin{equation}\label{coefref-B-circle}
    B=\frac{J_1\left( \sqrt{2 i t} r_+\right)}{J_1\left( \sqrt{2 i t} r_+\right) Y_0\left( \sqrt{2 i t} r_-\right)-J_0\left( \sqrt{2 i t} r_-\right) Y_1\left( \sqrt{2 i t} r_+\right)}.
\end{equation}
Then the average time needed to cross the absorbing boundary from Eq. \eqref{timetwoboundary} is obtained to be
\begin{equation}
\label{N-circleref}
   \big<\mathcal{N}\big>= \frac{1}{2} \left(-r^2+r_-^2+2 r_+^2 \log \Big(\frac{r}{r_-}\Big)\right) \, .
\end{equation}
One can easily check that in the limit where $r=r_-$ the average time goes 
to zero which makes sense since the field is immediately absorbed by the boundary. On the other hand, if one sets $r_-=0$ the average time goes to 
infinity since the field is never absorbed while it can be reflected for infinite times from the outer boundary.   Moreover it is straightforward to check that the maximum value of the above expression for a given $r_-$ occurs at $r=r_+$. This also makes sense, since inflation is terminated when the field hits the absorbing boundary and 
the maximum time occurs when the field starts from the farthest distance from the inner boundary which is $r_+$.  
Finally, for the configuration where the positions of the reflective and absorbing boundaries are switched, the result is obtained  by replacing $r_- \leftrightarrow  r_+$ in the above expression. 

Having calculated $\langle\mathcal{N}\rangle$ for both cases of fully absorbing and the mixed absorbing-reflective boundaries, given respectively by Eqs. (\ref{timetwoboundary2}) and (\ref{N-circleref}), we can compare them with each other.  Suppose the initial configuration is the fully absorbing boundaries with  $\langle\mathcal{N}\rangle$ given by Eq. (\ref{timetwoboundary2}). Now suppose, with the same initial position of the field $r$, we change one boundary, say $r_+$,  to be reflective. On the physical grounds we expect that $\langle\mathcal{N}\rangle$ to increase compared to the original configuration. The reason is that as one boundary becomes reflective, we need the field to hit the remaining absorbing 
boundary  in order for inflation to end. However,  in the initial configuration inflation is terminated when the field hits either of the absorbing boundaries which will take less time. Indeed, with some algebraic efforts one can show that $\langle\mathcal{N}\rangle$  in Eq. (\ref{timetwoboundary2}) is always smaller than its counterpart given in Eq.  (\ref{N-circleref}). 

We comment that while we have calculated the first hitting probabilities $p_\pm$  to terminate inflation for the case of fully absorbing boundaries, but we did not calculate them
for the case of mixed boundaries. The reason is that when we have a reflective boundary, say at $r_+$, then no matter how many times the field hits the reflective boundary at $r_+$, inflation ends only when the field hits the absorbing boundary at
$r_-$. In this view, we take $p_+=0$ and $p_-=1$. Of course, one can consider a more complicated question of asking how many times the field hit the reflective boundary before hitting the absorbing boundary to terminate inflation? This is an interesting (and difficult) mathematical question which is beyond the scope of our current investigation.  

Although  in the mixed boundary case inflation ends when the field hits the absorbing boundary and $p_{-}=1$, one can ask about the final position of the field on the absorbing boundary. In other words what is the probability that the field hits a particular point on the absorbing boundary? To study this interesting question, we start by the following formula \cite{Assadullahi:2016gkk}
\begin{equation}
    \nabla ^2 p= 0\,,
\end{equation}
where $p$ is the first crossing probability of a special section on $r_-$. In other words if we split the absorbing boundary as $A=D\cup\bar{D}$, where $D$ is the region of interest and $\bar{D}$ is its complement, by the boundary conditions, $p(D)=1$, $p(\bar{D})=0$ and $\nabla p.\hat{n}=0$ ($\hat{n}$ is the normal vector on the reflective boundary) one can solve the above equation for $p$. If we define $D=(\theta',\theta'+d\theta')$ then we obtain the probability density as
\begin{equation}
 f(\theta',\theta)=\frac{1}{2\pi}+ \sum _{m=1}^{\infty } \frac{(r_-^{2m}+r^{2m}) (\frac{r_+}{r})^m \cos \left(m \left(\theta' -\theta\right)\right)}{\pi(r_-^{2m}+r_+^{2m})} ,
\end{equation}
where $r$ and $\theta$ are the initial positions of the field and $-\pi+\theta<\theta'<\pi+\theta$.
An interesting property of the above equation is that in the limit we set $r=r_-=0$, the result behaves as a uniform density function as expected. Moreover, one can easily check that the above function has a maximum at $\theta'=\theta$ which means the probability becomes maximum for the  initial angular position of the field. On the other hand, it is minimum at the opposite direction $\theta'=\theta\pm\pi$ as expected.


\subsection{Generalization to $n$-dimension}
\label{n-dim}

One can generalize the above results  to the more general case where $n>2$ fields are under Brownian motion in dS background subject to two concentric $(n-1)$-dimensional spherical boundaries located at $r_-$ and $r_+$.  As before, we can work with the higher dimensional spherical coordinate in which the radial coordinate is given by $r^2= \sum_{i=1}^n \phi_i^2$.

As in two-dimensional case one can show that with the homogeneous and isotropic reflecting and/or absorbing boundary conditions the characteristic function depends only on $r$, satisfying
\begin{equation}\label{charnsphere}
   \frac{1}{r^{n-1}} \frac{\partial}{\partial r}\Big(r^{n-1}\frac{\partial\chi_\mathcal{N}(t,r)}{\partial r} \Big) = -2it\chi_\mathcal{N}(t,r) \, .
\end{equation}
The solution to the above equation is given by
\begin{equation}\label{char-nsphere}
    \chi_\mathcal{N}(t,r) = A r^{\frac{2-n}{2}} J_{\frac{n-2}{2}}\big(- \sqrt{2 i t} r \big)+B r^{\frac{2-n}{2}} Y_{\frac{n-2}{2}}\big(- \sqrt{2 i t} r \big),
\end{equation}
where  $A$ and $B$ are two constants that are determined from the boundary conditions while $J_{\frac{n-2}{2}}$ and $Y_{\frac{n-2}{2}}$ are the Bessel functions of the first and the second kind respectively.  

\begin{itemize}
  \item \textbf{Absorbing boundaries}
\end{itemize}

First we consider the case where both boundaries are absorbing. As discussed 
in two-dimensional case, the boundary conditions in this case are given by $\chi_\mathcal{N}(t,r_+)=\chi_\mathcal{N}(t,r_-)=1$, in which $A$ and $B$ 
are obtained to be
\begin{equation}
  A= \frac{r_+^{\frac{n-2}{2}} Y_{\frac{n-2}{2}}\left(- \sqrt{2i t} r_-\right)- r_-^{\frac{n-2}{2}} Y_{\frac{n-2}{2}}\left(- \sqrt{2i t} r_+\right)}{ J_{\frac{n-2}{2}}\left(- \sqrt{2i t} r_+\right) Y_{\frac{n-2}{2}}\left(-\sqrt{2} \sqrt{i t} r_-\right)-J_{\frac{n-2}{2}}\left(- \sqrt{2i t} r_-\right) Y_{\frac{n-2}{2}}\left(- \sqrt{2i t} r_+\right)} \, ,
\end{equation}
and
\begin{equation}
  B=\frac{ r_+^{\frac{n-2}{2}} J_{\frac{n-2}{2}}\left(- \sqrt{2 i t} r_-\right)- r_-^{\frac{n-2}{2}} J_{\frac{n-2}{2}}\left(- \sqrt{2 i t} r_+\right)}{ J_{\frac{n-2}{2}}\left(- \sqrt{2i t} r_-\right) Y_{\frac{n-2}{2}}\left(- \sqrt{2i t} r_+\right)-J_{\frac{n-2}{2}}\left(- \sqrt{2 i t} r_+\right) Y_{\frac{n-2}{2}}\left(- \sqrt{2 i t} r_-\right)} \, .
\end{equation}

Again it can be checked that $\chi(t=0,r)=1$ and  $P(\mathcal{N},r)$ is normalized. By taking the time derivative of the characteristic function and using Eq. (\ref{timetwoboundary}),  one can show that
\begin{equation}
\label{N-n}
  \big<\mathcal{N}\big>=  \frac{  \left(r_+^n-r_-^n\right)-\left(r_+^2-r_-^2\right) (\frac{r_-r_+}{r})^{n-2}}{n(r_+^{n-2}-r_-^{n-2})}-\frac{r^2}{n} \, .
\end{equation}
The case $n=2$ requires special care as both the numerator and the denominator 
vanish. However, one can check that in the limit $n=2$ we obtain the previous result Eq. (\ref{timetwoboundary2}). To see this, use the limiting approximation  
$X^n\simeq X^2 \big(1+ (n-2) \ln(X) \big)$ for $n\rightarrow 2$ and expand $\langle \cN \rangle$ in Eq. (\ref{N-n}) to leading order, which yields to Eq. (\ref{timetwoboundary2}). 

Although the expression (\ref{N-n}) has been obtained for a higher dimensional field space, it can also be applied to the case $n=1$. This corresponds to the case where two absorbing boundaries are located at the position $x_1=r_-$ and $x_2= r_+$ with the field initially located somewhere in between. Setting $n=1$ in Eq. (\ref{N-n}) we obtain
\ba
\label{N-n=1}
\big<\mathcal{N}\big>= (r- r_-) (r_+ - r) \quad \quad (n=1) \, .
\ea
This result agrees with the result obtained in \cite{Firouzjahi:2018vet}  in which 
it is shown that $\langle \cN \rangle$ grows with the product of the distance between the position of the field to each boundary.

In the limit where  $r_-=0$ in which  the inner boundary is at the origin of the field space $n\ge 2$, we obtain
\begin{equation}\label{nrmzero}
    \big<\mathcal{N}\big>=\frac{r_+^2-r^2}{n} \quad \quad (r_-=0, n\ge 2)      \, .
\end{equation}
If we further set the initial condition $r=0$, then we obtain $\big<\mathcal{N}\big>=\frac{r_+^2}{n}$. This is inline with our previous discussion that the average time to traverse the field space under Brownian motion is proportional to the square of the  distance  with the proportionality being $1/n$. 

\begin{figure}
        \centering
        \includegraphics[width=0.42\linewidth]{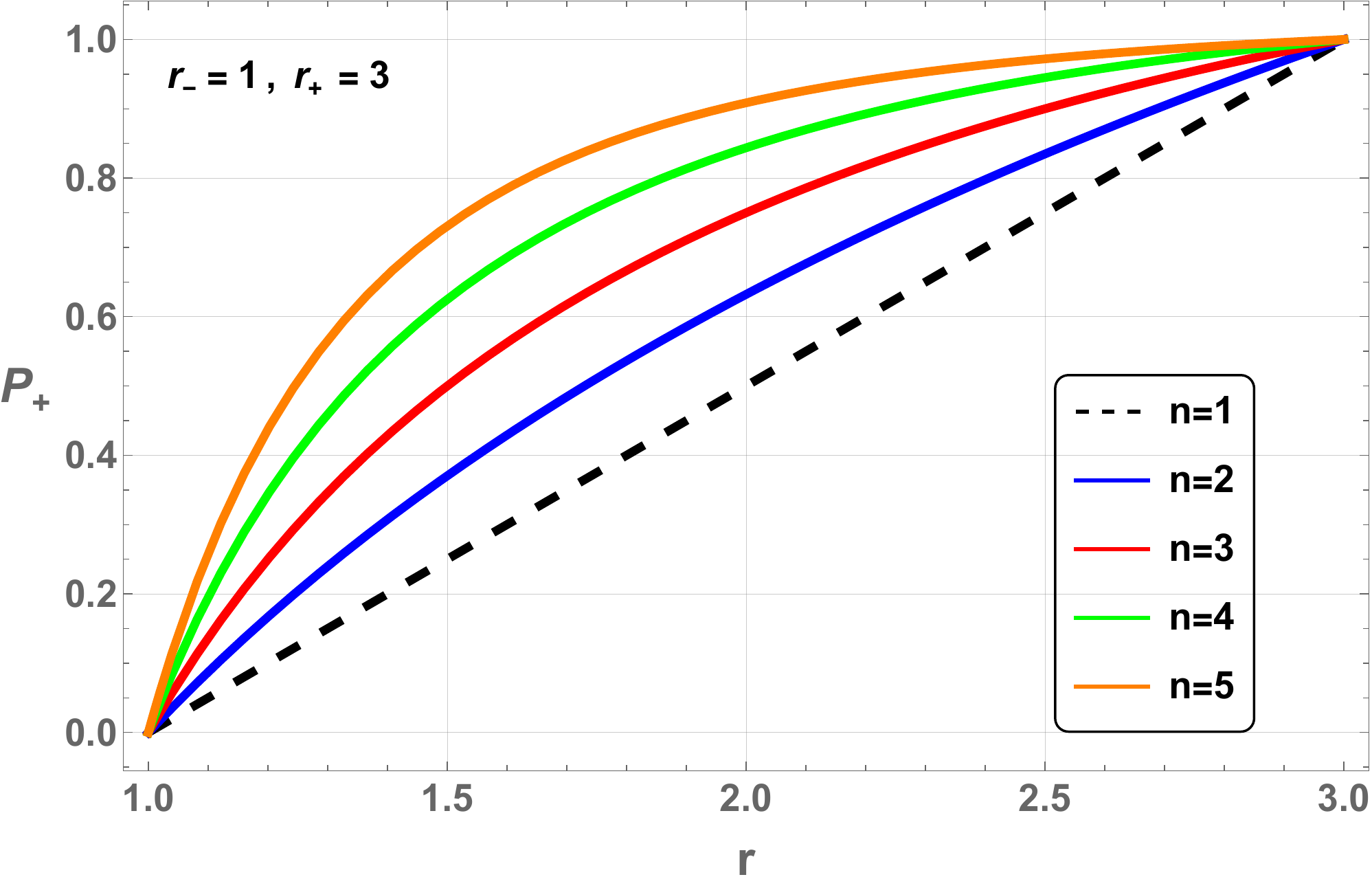}\hspace{2em}
        \includegraphics[width=0.4\linewidth]{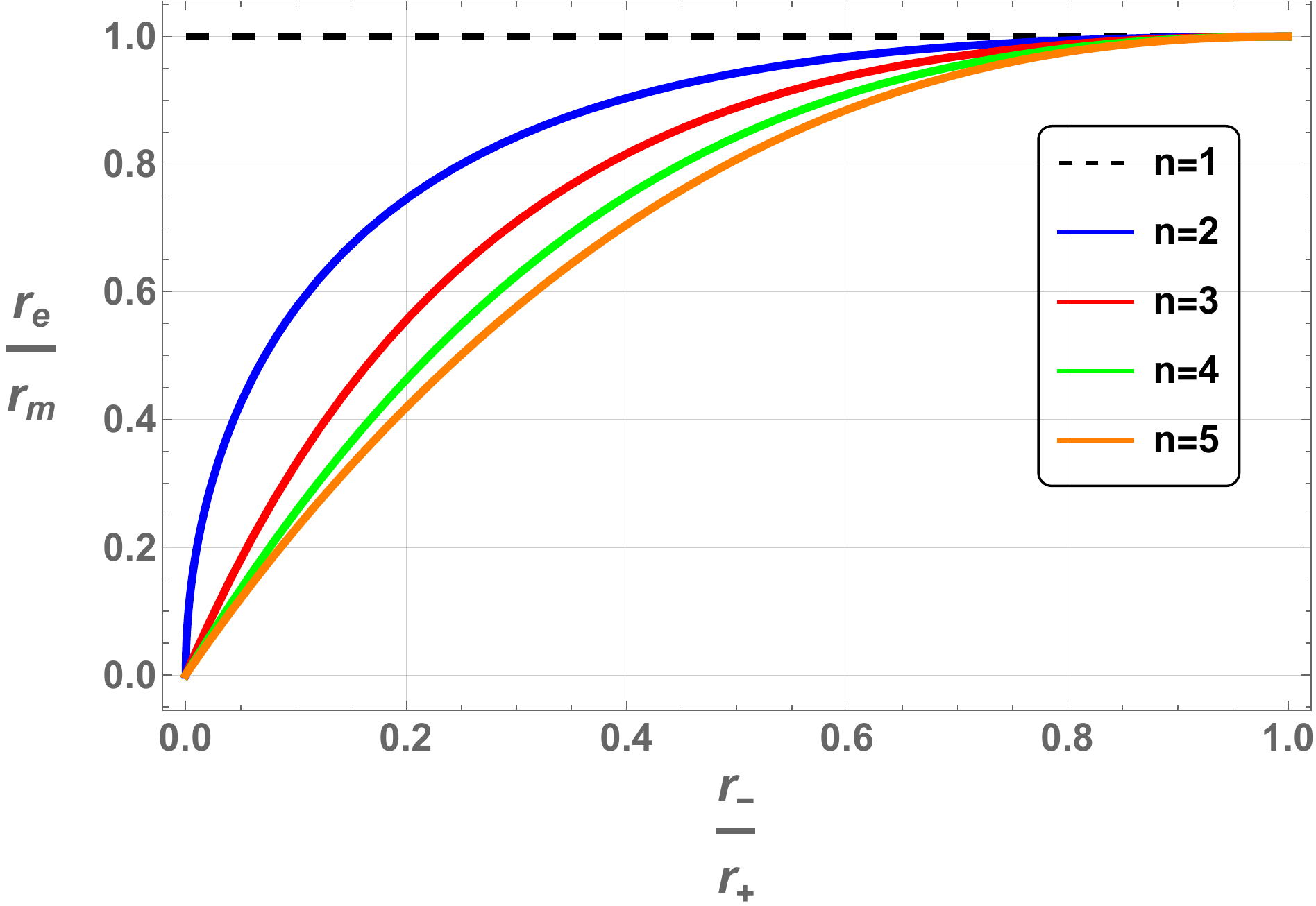}
    \caption{Left: the behaviour of $p_+$ as a function of $r$ for different dimensions. 
     $p_+$ grows linearly and logarithmically for $n=1$ and $n=2$ respectively, while its growth becomes milder for larger values of $n$. Right: the plot of $\frac{r_e}{r_m}$ as a function of $\frac{r_-}{r_+}$ in which $r_m=\frac{r_+ + r_-}{2}$ is the midpoint. For fixed values of $r_\pm$, we see that $r_e$ falls off as $n$ increases. This means that by increasing the dimensionality of field space, there is more volume for the field to wander around and it is more likely that the field hits the large exterior boundary than  the small interior boundary.}
    \label{probabilitysphere}
\end{figure}

Now one can calculate the first crossing probabilities $p_\pm$ as well. Defining the Wiener process for the integral solution of each field as in Eq. (\ref{sol-formal}) and
following the same steps  as  in the previous section, one can write
\begin{equation}\label{p+-n}
    p_+(r_+^2-r^2)+p_-(r_-^2-r^2)=n\big<\mathcal{N}\big> \, .
\end{equation}
Using the normalization condition $p_+ +  p_- =1$, this yields
\begin{equation}
\label{ppm}
\begin{split}
p_+=\frac{1- \big(\frac{r_{-}}{r} \big)^{n-2}}{ 1- \big(\frac{r_-}{r_+} \big)^{n-2}}\,
 \end{split}  \, , \quad \quad
 p_-=\frac{1- \big(\frac{r_{+}}{r} \big)^{n-2}}{ 1- \big(\frac{r_+}{r_-} \big)^{n-2}} \, .
\end{equation}
One can check that in the limit where $r_-=0$, we have $p_+=1$ which is the expected result as there is only the outer boundary for the field to hit. 
In the limit $n=1$, we obtain 
\ba
\label{p+n=1}
p_+ = \frac{r- r_-}{r_+ - r_-} \quad \quad (n=1) \, ,
\ea
so with fixed values of $r_-$ and $r_+$,   $p_+$ increases linearly with $r$ as obtained in 
\cite{Firouzjahi:2018vet}. For the limit $n=2$, we obtain our previous result 
Eq. (\ref{ppm2}), so $p_+$ increases only logarithmically with $r$. We expect that the growth of $p_+$ to become milder by increasing the dimensionality of field space. The behaviour of  $p_+$ for various values of the dimension of field space $n$ as a function of the initial position $r$ can be seen in Fig. \ref{probabilitysphere}. For all values of $n$, we have set the interior and the exterior radii to be fixed at $r_-=1$ and $r_+=3$ respectively.  As one expects, for a given value of $n$, $p_+$ is a monotonically increasing function of the initial position of the field $r$. 
 For starting points close to the inner boundary $p_+$ is close to zero. This makes sense because in this situation  it is easier for the field to hit the inner boundary 
 earlier than the outer boundary which is far away. The situation is reversed for starting points close to outer boundary. One non-trivial result is that for a given value of the initial position $r$, the probability $p_+$ increases by increasing $n$.
 Intuitively speaking, by increasing the dimensionality of the field space, there will be more volume for the field to wander around and the field tends to hit the large exterior boundary more frequently than to hit the inner boundary. 
 
 One interesting question is to calculate the  point of equal probability, i.e. the 
 initial position of the field, $r= r_{e}$, where $p_+=\frac{1}{2}$ so it is equally probable  for the field to hit either of the boundaries. Using Eq. (\ref{ppm}) we obtain \ba
 \label{r-half}
 r_{e} = r_-r_+ \Big[ \frac{2}{r_+ ^{n-2} +  r_- ^{n-2} } 
 \Big]^{\frac{1}{n-2}} \, .
 \ea 
In particular, for $n=1$ we obtain $r_{e} = \frac{r_- + r_+}{2}$ which is one expects, i.e. the equal probability point is the midpoint  between the two boundaries. On the other hand, for $n=2$, we obtain $r_{e}  = \sqrt{r_- r_+}$ which is the geometric average distance between the two boundaries. 
 The behaviour of $r_{e}$ for various values of $n$ as a function of the ratio 
 $\frac{r_-}{r_+}$ are plotted in right panel of Fig.  \ref{probabilitysphere}. 
 In this plot we have scaled  $r_e$ in units of the midpoint 
 $r_m \equiv \frac{r_-+ r_+}{2}$. As expected, for $n=1$, we have $r_e= r_m$.
 However, for a given ratio of $\frac{r_-}{r_+}$, the ratio $\frac{r_e}{r_m}$ falls off as $n$ increases.  This means that the point of equal probability approaches the interior boundary. In addition, for a given value of $n \neq 1$ and with a fixed value of  $r_+$, as $r_-$ decreases  the value of $\frac{r_e}{r_m}$ falls off as well. Both of the above conclusions are consistent with our previous results that by increasing the volume of the field space it is more likely that the field hits the large exterior boundary than the smaller interior boundary.

\vspace{0.5cm}
\begin{itemize}
  \item \textbf{Mixed boundaries}
\end{itemize}

Now suppose $r_+$ is the reflective boundary while $r_-$ is absorbing. 
In this case we have
\begin{equation}\label{coefref-A-nsphere}
    A=\frac{r_-^{\frac{n}{2}-1} Y_{\frac{n}{2}}\left((-1-i) \sqrt{t} r_+\right)}{ J_{\frac{n-2}{2}}\big((-1-i) \sqrt{t} r_-\big) Y_{\frac{n}{2}}\big((-1-i) \sqrt{t} r_+\big)- J_{\frac{n}{2}}\big((-1-i) \sqrt{t} r_+\big) Y_{\frac{n-2}{2}}\big((-1-i) \sqrt{t} r_-\big)} ,
\end{equation}
and
\begin{equation}\label{coefref-B-nsphere}
    B=\frac{ r_-^{\frac{n}{2}-1} J_{\frac{n}{2}}\left((-1-i) \sqrt{t} r_+\right)}{ J_{\frac{n}{2}}\left((-1-i) \sqrt{t} r_+\right) Y_{\frac{n-2}{2}}\left((-1-i) \sqrt{t} r_-\right)- J_{\frac{n-2}{2}}\left((-1-i) \sqrt{t} r_-\right) Y_{\frac{n}{2}}\left((-1-i) \sqrt{t} r_+\right)}.
\end{equation}

Having obtained the characteristic function and by taking its time derivative,
one can show that the average time is given by
\begin{equation}
\label{N-sphereref}
    \big<\mathcal{N}\big>=\frac{r_-^2-r^2}{n}+\frac{2r_+^n}{n(n-2)}(r_-^{2-n}-r^{2-n}) \, .
\end{equation}
Interestingly, one can check that in the limit of $n\rightarrow 2$, the above result reduces to our old formula in two dimension given in Eq. (\ref{N-circleref}). 
On the other hand, for the case $n=1$ the above equation reduces to
\begin{equation}
\label{N-n=1}    \big<\mathcal{N}\big>=(r-r_-)(2r_+-r-r_-) \quad \quad   (n=1) \, .
\end{equation}

As in the case of $n=2$, one can  show that with the fixed values of $r_\pm$ and $r$, the value of $\langle \cN \rangle$ given in Eq. (\ref{N-sphereref}) is larger than 
its corresponding value in the configuration of fully absorbing boundaries given in 
Eq. (\ref{N-n}).  The interpretation is the same as before: it is more frequent to terminate inflation by hitting either of two absorbing boundaries than hitting only one absorbing boundary. Finally, as in the case of $n=2$, we do not calculate the first hitting probabilities $p_\pm$ to terminate inflation. Since one boundary is reflective, by construction inflation is terminated only when the field hits the absorbing boundary, yielding the trivial result $p_\pm =0$ or $1$, depending on the relative configurations of the absorbing/reflective boundaries. 


\subsection{Power Spectrum }

Using the stochastic $\delta N$ formalism \cite{Fujita:2013cna, Fujita:2014tja, Vennin:2015hra, Vennin:2016wnk}, we can define the curvature perturbations associated to fields perturbation via $\calR = -\delta \cN$. There is no classical rolling of the scalar fields as in conventional slow-roll models. However, the existence of boundaries allows  a well-defined realization of the number of e-folds $\cN$ which we take to carry the same meaning as in models of inflation dominated by the classical drift \cite{Vennin:2016wnk}.

In the case that one of the barriers is reflective we have a well defined surface of energy for the motion of the field and we can use stochastic $\delta N$ formalism to calculate the power spectrum. The power spectrum of curvature perturbation is related to the variance $\langle\delta \cN^2\rangle \equiv 
\langle \cN^2 \rangle - \langle \cN \rangle^2$ via \cite{Vennin:2015hra, Vennin:2016wnk}
\begin{equation}
\label{power}
    \mathcal{P}_\calR=\frac{d\langle \delta\mathcal{N}^2\rangle}{d\left<\mathcal{N}\right>} \, ,
\end{equation}
in which $\langle \cN^2 \rangle $ is calculated using Eq. (\ref{timetwoboundary})
setting $s=2$.  

To calculate the power spectrum one has to specify the trajectory in field space. 
In our case at hand, since the boundaries are symmetric, the only relevant variable
is the radial distance $r$ of the initial position in field space  so we can treat 
$ \cN$ and $\delta \cN$ as functions of $r$ and perform the differentiations.  

While we have presented the expression for $ \langle \cN \rangle$ in previous subsections, we do not write the corresponding formula for $\delta \cN$ as the result is complicated. Instead, we present the final result for the curvature perturbation power spectrum in 
a $n$-dimensional field space. After restoring the factors of  $\frac{H}{2 \pi}$, the power spectrum in a configuration in which $r_+$ is reflective and $r_-$ is absorbing is given by 
\ba
\label{power_sphere}
   \mathcal{P_\calR}= \frac{16 \pi^2 \Big[   r_+^n \left(n^2 r_+^2-\left(n^2-4\right) r^2\right)+r^2 \Big( (n-2) r^{ n}-(n+2)  r^{-n} r_+^{2 n}\Big) \Big] }{n \left(n^2-4\right) \big(r^n-r_+^n\big) H^2}  \,  .
   \ea
Curiously, we see that the power spectrum is independent of $r_-$, the position of the absorbing boundary, where inflation is terminated. It only depends on the initial position $r$ and the position of the outer reflective boundary. 
 
The power spectrum for the spacial case $n=2$ where Eq. (\ref{power_sphere}) 
 is seemingly singular can be obtained from its leading expansion, yielding  
\ba
\label{power-n=2}
 \mathcal{P_\calR} = \frac{2 \pi^2}{ H^2} \Big[ \frac{4 r_+^4}{ r_+^2- r^2} \ln \Big(\frac{r_+}{r} \Big) - \Big( 3 r_+^2 - r^2 \Big)  \Big] \, \quad \quad (n=2) \, .
\ea  
The power spectrum for the switched boundaries in which $r_-$ is reflective while $r_+$ is absorbing is simply given by replacing $r_- \leftrightarrow r_+$ in the above expressions. 

\begin{figure}
    \centering
    \begin{subfigure}[t]{0.48\textwidth}
        \centering
        \includegraphics[width=\linewidth]{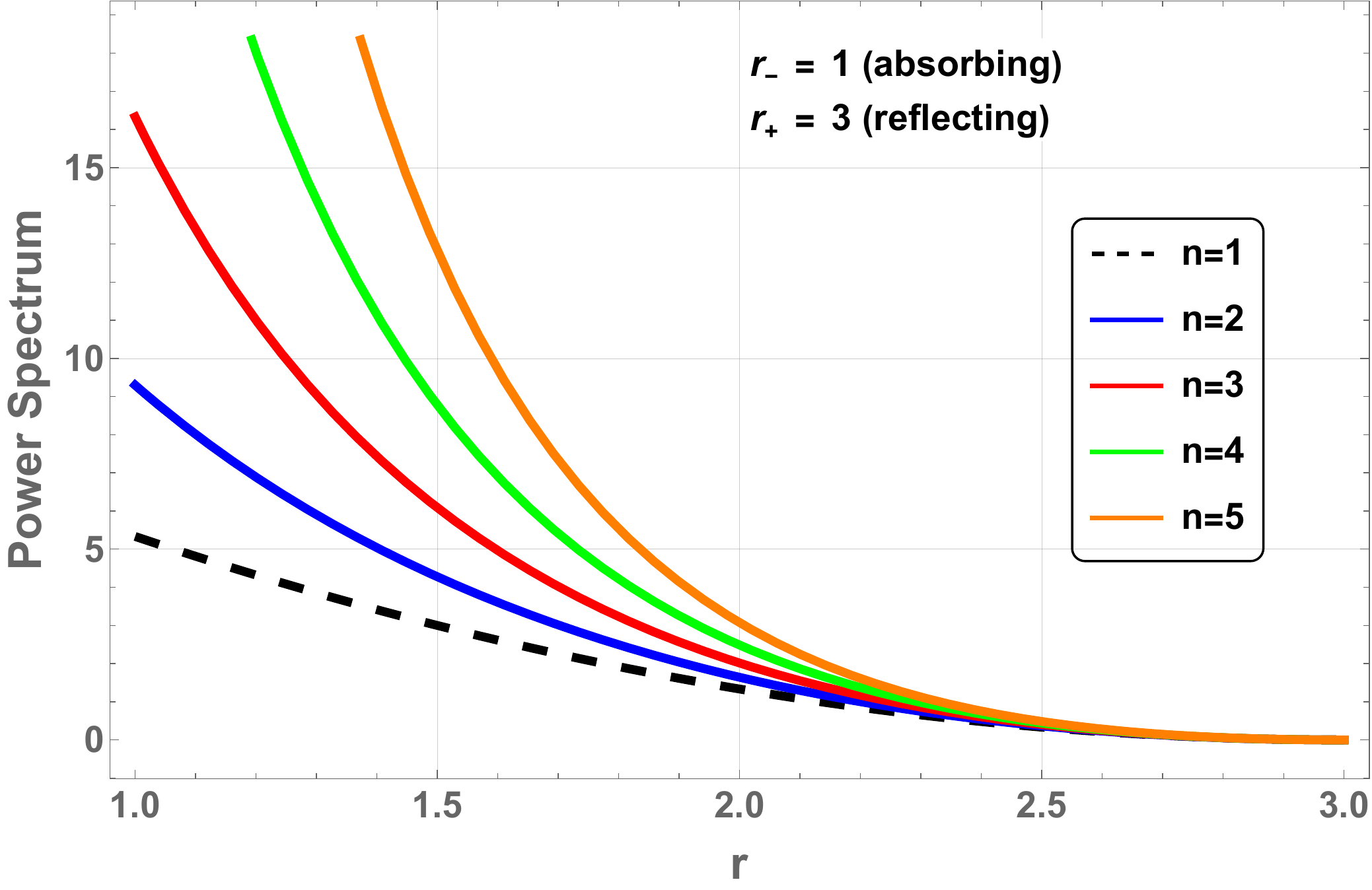}
    \end{subfigure}
    \hfill
    \begin{subfigure}[t]{0.47\textwidth}
        \centering
        \includegraphics[width=\linewidth]{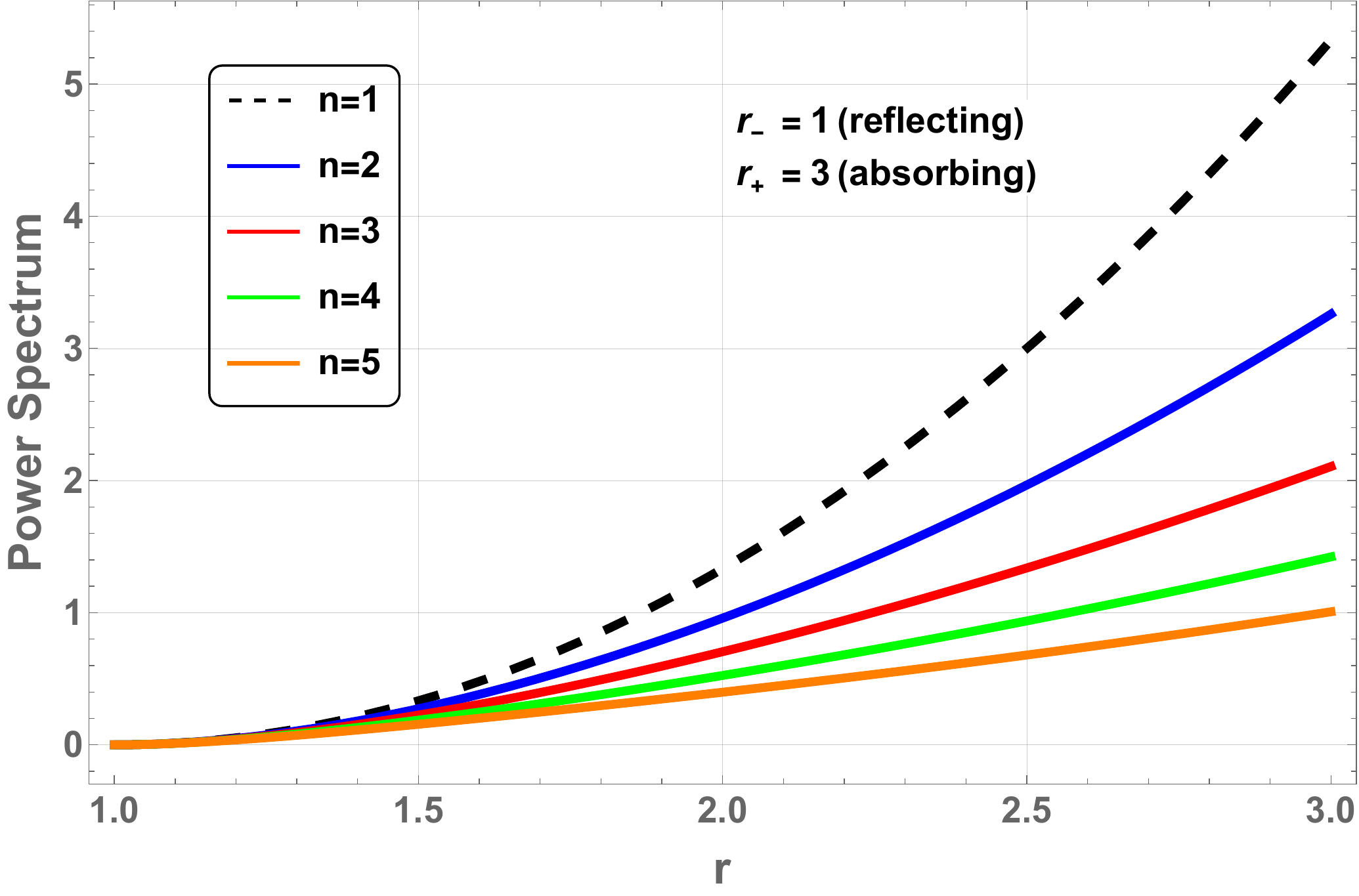}
    \end{subfigure}
    \caption{The behaviour of power spectrum versus $r$ for various dimensions and for two different boundary conditions. In the left panel the inner boundary is absorbing while the outer one is reflective. As the dimension of field space increases the power spectrum increases as well.  In the right panel the position of the reflective and absorbing boundaries are switched in which we see the opposite trend compared to the left panel.
    }
  \label{power_Circle}
\end{figure}

Fig. \ref{power_Circle} shows the behaviour of power spectrum versus the initial position of the field $r$ in various dimensions for different configurations of the  boundaries. In the left panel the interior boundary 
$r_-=1$ is absorbing while the outer boundary $r_+=3$ is reflective. As we see, 
for a given initial condition $r$, by increasing the dimension of the field space the power spectrum increases as well. In the right panel the position of the reflective and absorbing boundaries are switched  where now we see a reverse trend in which by increasing the dimension of the field space the power spectrum decreases. This behaviour may be interpreted as follows. The surface of end of inflation (or end of USR phase) is given by the absorbing boundary. Intuitively, one expects that the power spectrum increases when the field has less chance to hit the absorbing boundary. As we discussed previously, this happens when the outer boundary is reflective and the inner boundary is absorbing (as in left panel of Fig. \ref{power_Circle} ). In this configuration,  by increasing $n$ the fields spend more time near the large outer boundary than the small interior boundary.

\section{Asymmetric Boundaries}
\label{asymmetry}
After studying the setups with symmetric boundaries in two and higher dimensions, now we study the  cases where the boundaries are asymmetric. 
Since we lose the symmetry, the analyses are complicated, so we restrict ourselves to two-dimensional field space. We study  two shapes of the boundaries, rectangle and  sector and leave the study for more complicated boundaries such as  
hyperbola and ellipsoid to future studies.   

\subsection{Rectangle Boundary}
\label{rectangle}

\begin{figure}
 \vspace{-1cm}
    \centering
   \includegraphics[scale=0.55]{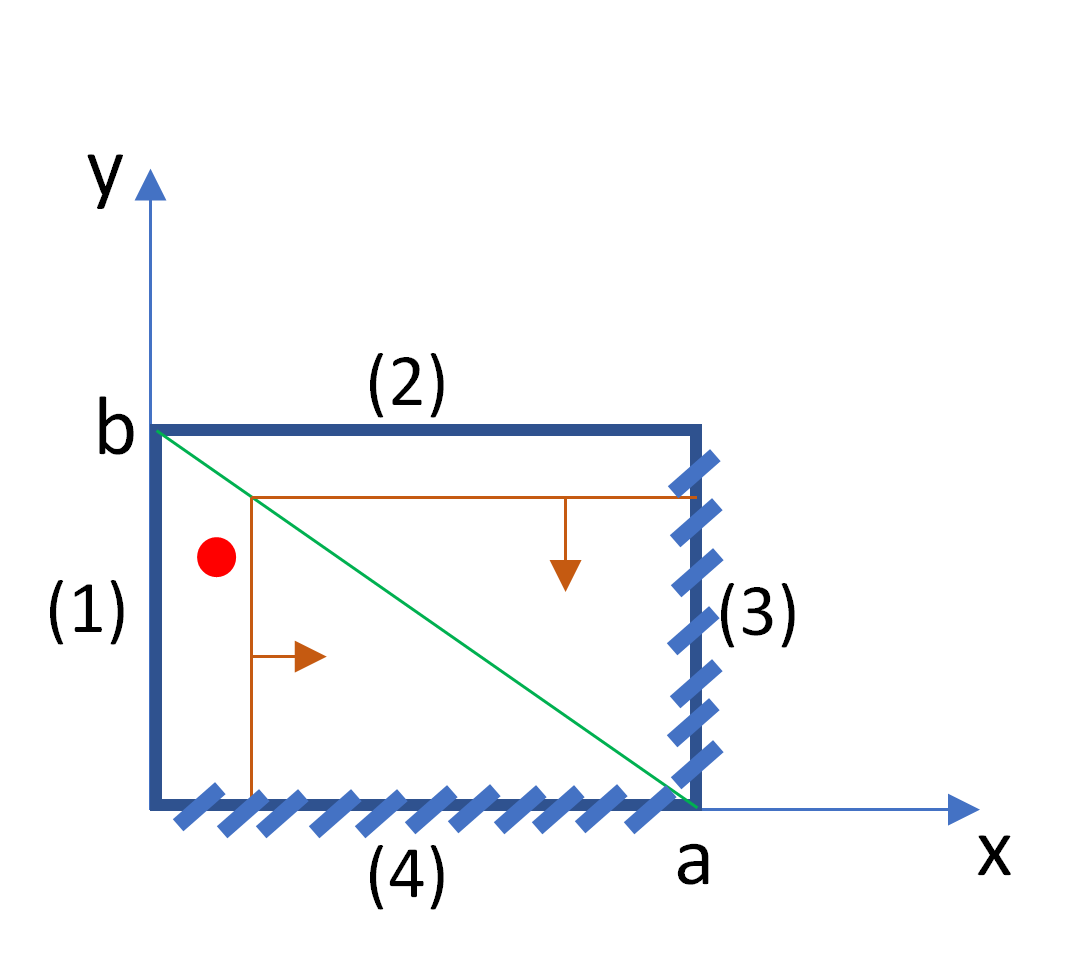}
    \caption{A schematic view of the rectangle boundary.  In this figure sides $(3)$ and $(4)$  are reflective while the other two sides are absorbing. The red lines and arrows show the directions of derivatives in calculating the power spectrum according to the region where the field is located.}
    \label{Boundariesrec}
\end{figure}

First we consider the configuration where the boundaries are given by a rectangle.
Various boundary conditions can be imagined for each side of the rectangle. As some sample studies we consider two cases in detail: case 1 where all the sides are absorbing boundaries and case 2 where two sides are reflective while the other two sides are absorbing. Fig. \ref{Boundariesrec} illustrates a schematic view of this boundary in which sides number $(3)$ and $(4)$  are  reflective while the other two sides are absorbing.

Using the adjoint Fokker-Planck equation one has
\begin{equation}
    \frac{\partial P(\mathcal{N},x,y)}{\partial \mathcal{N}}=\frac{1}{2}\frac{\partial^2 P(\mathcal{N},x,y)}{\partial x^2}+ \frac{1}{2}\frac{\partial^2 P(\mathcal{N},x,y)}{\partial y^2}.
\end{equation}
Defining $S_i$ as the $i$-th side of the rectangle, the absorbing boundary condition  is $P(S_i,\mathcal{N})=\delta(\mathcal{N})$ while for 
the  reflective boundary it  is $\nabla P(S_i,\mathcal{N}).\hat{n}=0$ where $\hat{n}$ is the unit orthogonal vector to the reflective boundary. 

As before,  to use the method of characteristic function one needs the Fourier transformation of the above adjoint Fokker-Planck equation which reads
\begin{equation}\label{FP-rec}
    \frac{\partial^2 \chi_\mathcal{N}}{\partial x^2}+ \frac{\partial^2 \chi_\mathcal{N}}{\partial y^2}=-2it \chi_\mathcal{N}\,,
\end{equation}
with the condition $\chi_{\mathcal{N}}(S_i,\mathcal{N})=1$ for the absorbing and $\nabla \chi_\cN(S_i,\mathcal{N}).\hat{n}=0$ for reflective boundaries.

\begin{itemize}
  \item \textbf{Case 1: all absorbing boundaries}
\end{itemize}

Since all the sides are considered to be absorbing, the boundary conditions are 
as follows,  
\begin{equation}
    \chi_\mathcal{N}(t,0,y)=\chi_\mathcal{N}(t,a,y)=\chi_\mathcal{N}(t, x,0)=\chi_\mathcal{N}(t,x,b)=1 \, ,
\end{equation}
in which $a$ and $b$ are the lengths of the two sides of the rectangle. 

In order to solve Eq. \eqref{FP-rec} with the mentioned boundary conditions we decompose the problem into two  steps 
\begin{equation}\label{bound-rec-2}
    \chi_{ \mathcal{N}1}(t,0,y)=\chi_{ \mathcal{N}1}(t,a,y)=1 ,\,\, \chi_{ \mathcal{N}1}(t,x,b)=\chi_{ \mathcal{N}1}(t,x,0)=0,
\end{equation}
and
\begin{equation}
    \chi_{ \mathcal{N}2}(t,x,b)=\chi_{ \mathcal{N}2}(t,x,0)=1 ,\, \, \chi_{ \mathcal{N}2}(t,0,y)=\chi_{ \mathcal{N}2}(t,a,y)=0.
\end{equation}
Finally after obtaining $\chi_{ \mathcal{N}1}$ and $\chi_{ \mathcal{N}2}$ one can write the general result as $\chi_{ \mathcal{N}}=\chi_{ \mathcal{N}1}+\chi_{ \mathcal{N}2}$.

Solving Eq. \eqref{FP-rec} for $\chi_{ \mathcal{N}1}$  we obtain
\begin{equation}
    \chi_{ \mathcal{N}1}(t,x,y)= \Big[ A_k \sin(k y)+B_k \cos(ky) \Big]\Big[C_k \sin\big(\sqrt{2it-k^2}x\big)+C_k \cos\big(\sqrt{2it-k^2}x \big)\Big].
\end{equation}
Imposing the boundary conditions \eqref{bound-rec-2} to calculate the coefficients, 
$ \chi_{ \mathcal{N}1}(t,x,y)$ is obtained to be 
\begin{equation}
    \chi_{\mathcal{N}1}(t,x,y)= \sum_{k=0}^{\infty}{\Big[a_k \cos(\alpha_k x)+b_k \sin(\alpha_k x)\Big]}
    \sin\big(\frac{k \pi}{b}y \big) \,,
\end{equation}
where, 
\begin{equation}
    a_k= \frac{2(1-(-1)^k)}{k\pi}\,\,,\,\,
    b_k= \tan \Big[\frac{a}{2}\sqrt{2i t-(\frac{k \pi}{b} )^2}~ \Big]a_k\,,
\end{equation}
and
\begin{equation}
    \alpha_k=\sqrt{2it-\big(\frac{k \pi}{b} \big)^2}\,.
\end{equation}
As we see from the form of $a_k$, only the odd modes contribute into the sum, $k= 2n+1$ with $n=0, 1, ... .$.

Following the same procedure for $\chi_{\mathcal{N}2}$,  the characteristic function $\chi_{\mathcal{N}}= \chi_{\mathcal{N}_1}+ \chi_{\mathcal{N}_2}$ is obtained as follows
\begin{equation}
    \chi_{\mathcal{N}}(t,x,y)= \sum_{k=0}^{\infty}  \Big\{ \big[a_k \cos(\alpha_k x)+b_k \sin(\alpha_k x ) \big] \sin\big(\frac{k \pi}{b}y\big)+ \big[{a}_k \cos(\gamma_k y)+\tilde{b}_k \sin(\gamma_k y)\big] 
    \sin \big(\frac{k \pi}{a}x \big) \Big\}\,,
\end{equation}
where
\begin{equation}
    \gamma_k=\sqrt{2it-\big(\frac{k \pi}{a}\big)^2}\, , \quad 
    \tilde{b}_k=  \tan \Big[\frac{b}{2}\sqrt{2i t-(\frac{k \pi}{a})^2} ~ \Big] {a}_k\, .
\end{equation}

Having calculated  $\chi_{\mathcal{N}}$, the average  $\left\langle \mathcal{N}\right \rangle$ can be obtained via the following relation,
\begin{equation}
\label{N-av}
    \left<\mathcal{N}\right>=-i\frac{d\chi(t)}{dt}|_{t=0} \, .
\end{equation}

One can also proceed and calculate the probability of hitting one edge of the rectangle before the other three ones \cite{Assadullahi:2016gkk}. As it is shown in \cite{Assadullahi:2016gkk}, the probability of crossing the $i$-th barrier before crossing the other ones satisfies the following equation, 
\begin{equation}
    \nabla^2p_i=0,
\end{equation}
with   $i=1,2,3,4$ while we set $p_i=\delta_{ij}$ on the $j$-th barrier.  Without loss of generality we take $i=2$  which is set at the edge $y=b$ of the rectangle. Using the method of the separation of variables we obtain
\begin{equation}
    p_2=\sum_{k=1}^{\infty}\left[ A_k \sinh\big(\frac{k \pi }{b} y \big)+B_k \cosh\big(\frac{k \pi }{b} y \big)\right] \sin\big(\frac{k \pi x}{a}\big).
\end{equation}
Since on $y=0$ the probability vanishes, we obtain $B_k=0$. Imposing the other boundary condition at $y=b$, yields 
\begin{equation}
    A_k=\frac{2(1-(-1)^k)}{k \pi \sinh(k \pi)}.
\end{equation}
The other probabilities $p_{2,3,4}$ can be obtained in a similar way which for brevity  we don't present the results here.

\begin{figure}
    \centering
    \begin{subfigure}[t]{0.48\textwidth}
        \centering
        \includegraphics[width=\linewidth]{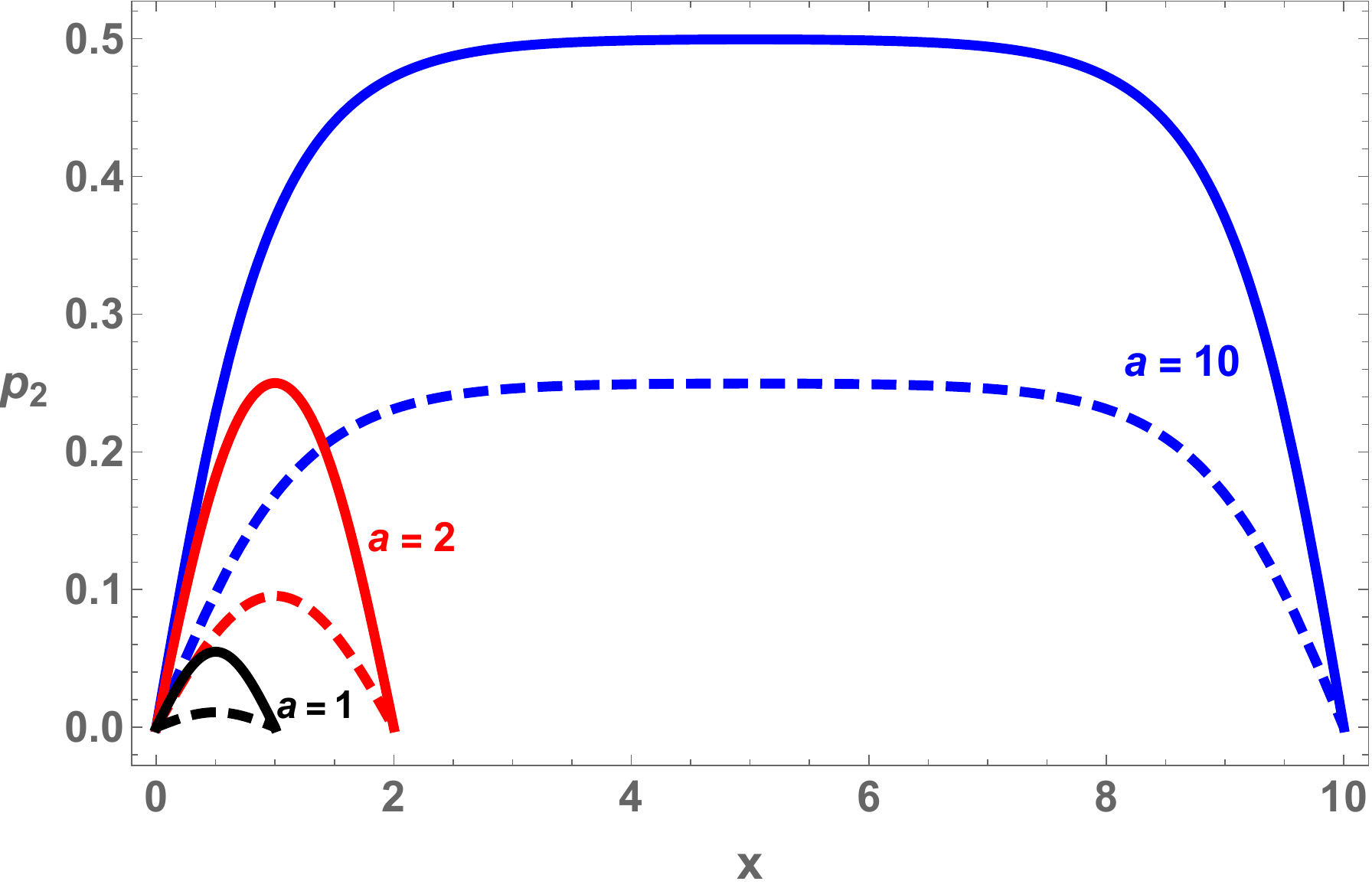}
    \end{subfigure}
     \caption{The probability of crossing the upper horizontal side  of the rectangle 
     before hitting other sides versus $x$ for different values of $a$ but with fixed $b=2$.  Each solid curve represents the case with the initial position  $y=\frac{b}{2}$ while the dashed curve with the same color  represents the case with the 
     initial position $y=\frac{b}{4}$. As the ratio $\frac{a}{b}$ becomes very large, the system effectively approaches  a one-dimensional configuration and the upper two curves approach respectively to 
 $p_+= \frac{1}{2}$ and $p_+= \frac{1}{4}$. 
  }
    \label{p2rectangleabsorbing}
\end{figure}

In Fig \ref{p2rectangleabsorbing} the probability $p_2$ of crossing the upper horizontal side (side 2) of rectangle  for various ratio of $\frac{a}{b}$ is plotted. In this plot we fix the height $b=2$ while varying the width $a$. In the limit  
$\frac{a}{b} \gg 1$, the system approaches a one-dimensional configuration 
and the expression for $p_2$ is given by Eq. (\ref{p+n=1}). For example, if the field starts with the initial configuration $y= \frac{b}{2}$, then $p_2$ approaches 
$\frac{1}{2}$ while if it starts from the initial position $y= \frac{b}{4}$, the probability $p_2$ approaches $\frac{1}{4}$. Both of these results can be obtained from Eq. (\ref{p+n=1}) as well. 

\begin{itemize}
  \item \textbf{Case 2: mixed boundary conditions}
\end{itemize}

Now we consider the case in which only two sides of the rectangle  are absorbing while the other two are reflective. We choose the sides $(x,y=0)$ and $(x=a,y)$ to be reflective, see Fig. \ref{Boundariesrec}.  
Correspondingly,  the boundary conditions for the characteristic function are
\begin{equation}\label{bound-rec-3}
    \chi_{ \mathcal{N}}(t,0,y)=\chi_{ \mathcal{N}}(t,x,b)=1 ,\,\,\,\,\,\,\, \frac{\partial\chi_{ \mathcal{N}}}{\partial x}\bigg|_{x=a}=\frac{\partial\chi_{ \mathcal{N}}}{\partial y}\bigg|_{y=0}=0 \, .
\end{equation}

\begin{figure}
    \centering
    \begin{subfigure}[t]{0.48\textwidth}
        \centering
        \includegraphics[width=\linewidth]{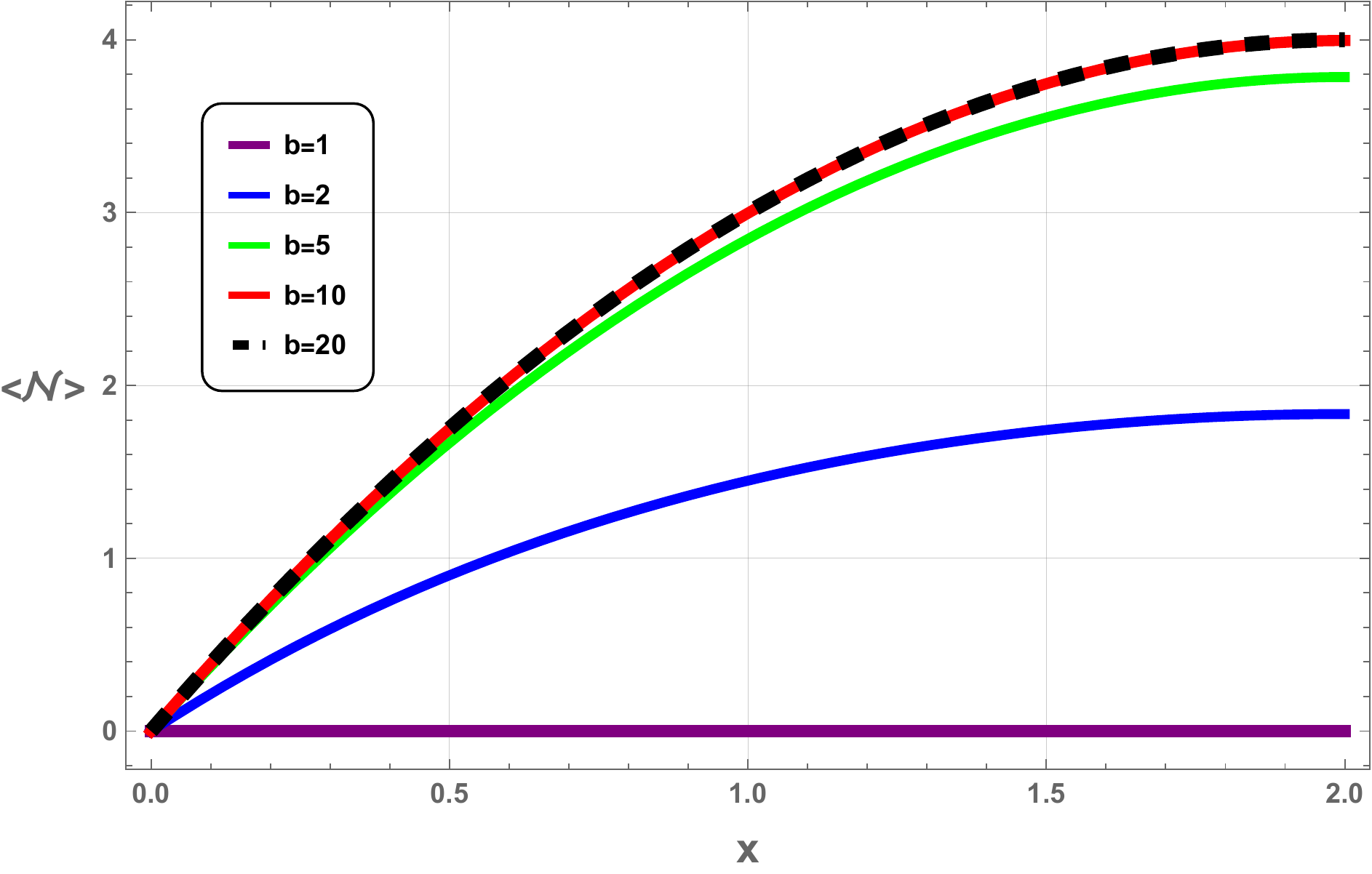}
    \end{subfigure}
    \hfill
    \begin{subfigure}[t]{0.46\textwidth}
        \centering
        \includegraphics[width=\linewidth]{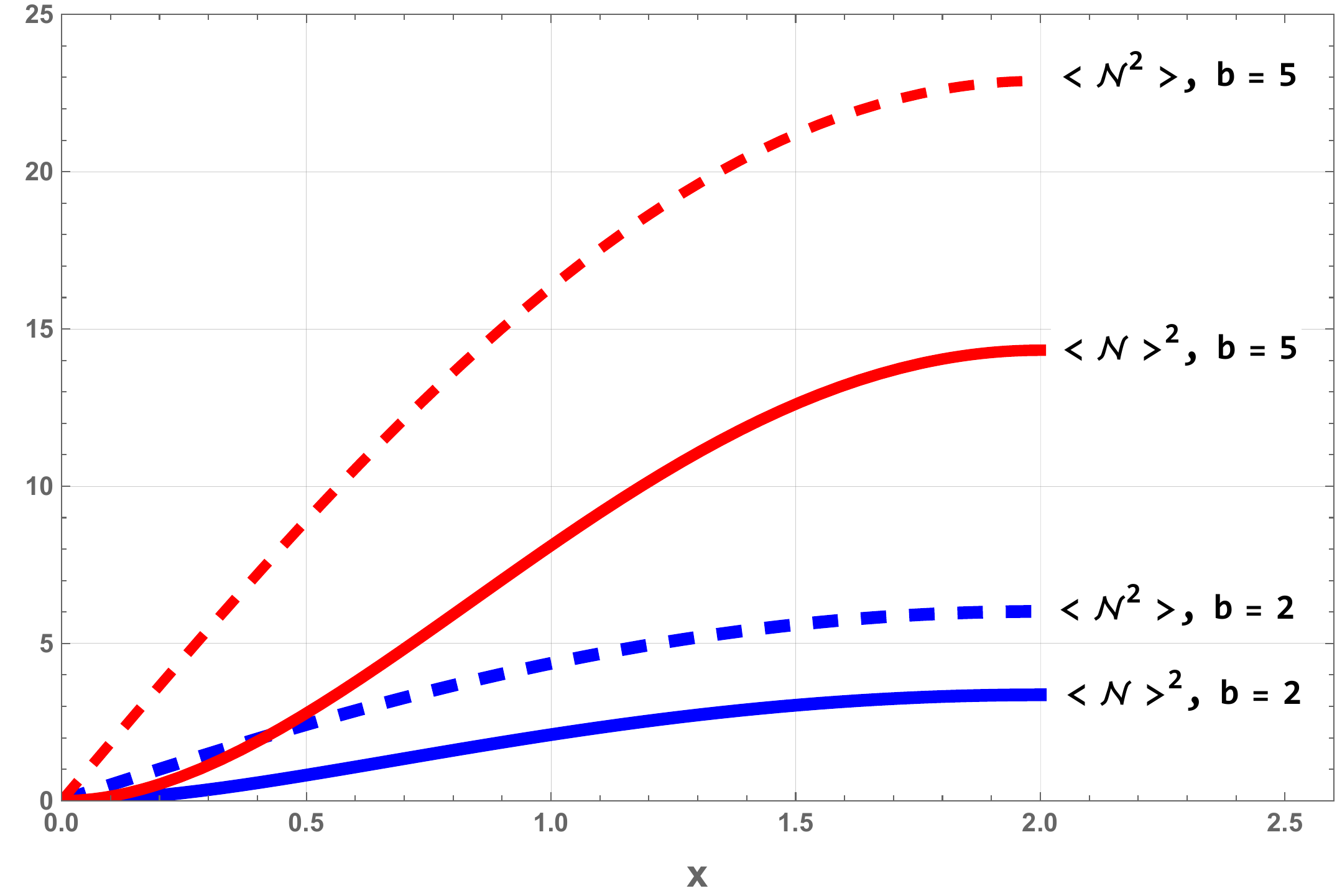}
    \end{subfigure}
    \caption{ Left: The behaviour of $\left<\mathcal{N}\right>$ in a rectangle with mixed boundary condition for fixed values of $a=2$ and $y=1$ but 
 different values of the height $b$. In the limit $\frac{b}{a} \gg1$ the result approaches to the one-dimensional configuration given by Eq. (\ref{N-n=1}). 
Right: The comparison of $\left<\mathcal{N}\right>$ and   $\left<\mathcal{N}^2\right>$  for two different values of $b$,  confirming that in each case 
  $\left<\mathcal{N}\right>^2 < \left<\mathcal{N}^2\right>$. 
    }
    \label{rectaglereflectiveaverage}
\end{figure}

To find the solution, we follow the same method as in the previous case  and decompose $ \chi_{ \mathcal{N}}$  into two parts. The first part is
\begin{equation}\label{bound-rec-5}
    \chi_{ \mathcal{N}1}(t,x,b)=1 ,\,\,\,\,\,\,\,\chi_{ \mathcal{N}1}(t,0,y)=0,\,\,\,\,\, \frac{\partial\chi_{ \mathcal{N}1}}{\partial x}\bigg|_{x=a}=\frac{\partial\chi_{ \mathcal{N}1}}{\partial y}\bigg|_{y=0}=0.
\end{equation}
Imposing the above boundary conditions on $\chi_{\mathcal{N}1}$ yields, 
\begin{equation}
    \chi_{\mathcal{N}1}=\sum_{k=0}^{\infty} {A_k \cos{(\gamma_k y)}\sin{\big(\frac{(2k+1)\pi}{2a} x\big)}}
\end{equation}
where
\begin{equation}
    A_k=\frac{4}{\pi(2k+1)\cos{(\gamma_k b)}} , \,\,\,\,\text{and}\,\,\,\,\,
    \gamma_k=\sqrt{2it-\big(\frac{(2k+1)\pi}{2a}\big)^2}.
\end{equation}
Now considering the second part, $\chi_{\mathcal{N}2}$, with the following boundary conditions, 
\begin{equation}\label{bound-rec-4}
    \chi_{ \mathcal{N}2}(t,0,y)=1 ,\,\,\,\,\,\,\,\chi_{ \mathcal{N}2}(t,x,b)=0,\,\,\,\,\, \frac{\partial\chi_{ \mathcal{N}2}}{\partial x}\bigg|_{x=a}=\frac{\partial\chi_{ \mathcal{N}2}}{\partial y}\bigg|_{y=0}=0,
\end{equation}
we obtain the following result
\begin{equation}
    \chi_{\mathcal{N}2}=\sum_{k=0}^{\infty} {\Big[ C_k \cos{(\tilde{\gamma_k} x)+D_k\sin{(\tilde{\gamma_k}x)}}\Big] \cos{\Big(\frac{(2k+1)\pi}{2b} y\Big)}} \, ,
\end{equation}
where
\begin{equation}
    C_k=\frac{4\cos{k\pi}}{\pi(2k+1)}\,,\,\,\,\,D_k=C_k \tan{(\tilde{\gamma_k} a)},\,\,\,
    \tilde{\gamma_k}=\sqrt{2it-\big(\frac{(2k+1)\pi}{2b}\big)^2}.
\end{equation}
Finally the  total characteristic function with the mixed absorbing and reflective sides is given by, 
 \begin{equation}
     \chi_{\mathcal{N}}=\chi_{\mathcal{N}1}+\chi_{\mathcal{N}2}.
 \end{equation}
 With the characteristic function  $\chi_{\mathcal{N}}$ at hand, $\left\langle \mathcal{N}\right \rangle$ is obtained as before via Eq. (\ref{N-av}). 
 In Fig. \ref{rectaglereflectiveaverage} we have compared the behaviour of $\left<\mathcal{N}\right>$ and $ \left<\mathcal{N}^2\right>$  versus $x$
 for  fixed values of $a$ and $y$
 but for  different values of the height $b$.  We see that as the ratio $\frac{b}{a}$ becomes very large the system effectively approaches a one-dimensional configuration and $\langle \cN \rangle$ approaches the value given by 
Eq.  \eqref{N-n=1}. 

Since we have reflective boundaries 3 and 4, we can not define the first hitting probabilities $p_3$ and $p_4$ to terminate inflation. However, since the boundaries 1 and 2 are absorbing, we can define the first hitting probabilities $p_1$ and $p_2$ to terminate inflation. As we have calculated these quantities in previous case of fully absorbing boundaries, here we do not present them for brevity.

 \begin{figure}
    \centering
    \begin{subfigure}[t]{0.48\textwidth}
        \centering
        \includegraphics[width=\linewidth]{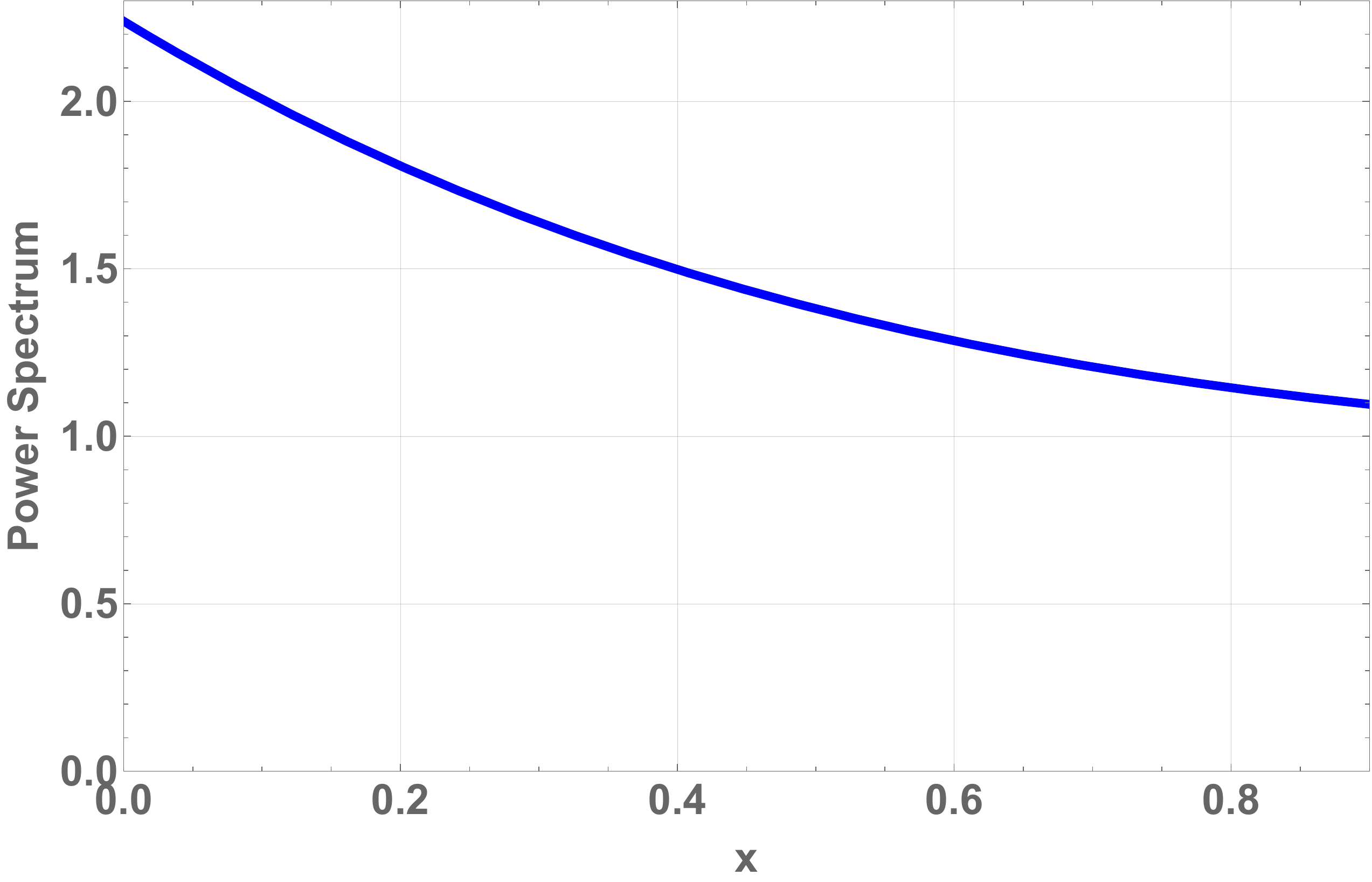}
    \end{subfigure}
     \caption{The power spectrum in the rectangle with two reflective sides (see Fig. \ref{Boundariesrec}) in units of $\frac{H}{2\pi}$ with $a=b=2$ and the initial condition $y=1$. In the 
     region $y<2-x$ the power spectrum is calculated orthogonal to the surface of end of inflation, i.e. derivative being along the $x$ direction. The results for $y>2-x$ are the same by switching $y\longleftrightarrow x$  for all steps.}
    \label{power-rectangle}
\end{figure}


 \subsubsection{Power Spectrum}
 The power spectrum  in this configuration is plotted in Fig \ref{power-rectangle} in units of $\frac{H}{2\pi}$ with $a=b=2$.  As the rectangle is symmetric with respect to the line $y=2-x$, we have calculated the power spectrum in the region $y<2-x$ with the initial condition $y=1$. In this region the derivatives to calculate the
power spectrum is along the direction orthogonal to the surface of end of inflation, 
i.e. derivative being along the $x$ direction.  The power spectrum in the upper region can be  calculated orthogonal to $y=b$ and the results don't change by  reflection to the line $y=2-x$. The red lines and arrows in Fig. \ref{Boundariesrec}  show the directions of derivatives in calculating the power spectrum according to the region where the field is located.

\subsection{Sector Boundary}\label{Section_Sector}

\begin{figure}
 \vspace{-0.5cm}
    \centering
   \includegraphics[scale=0.55]{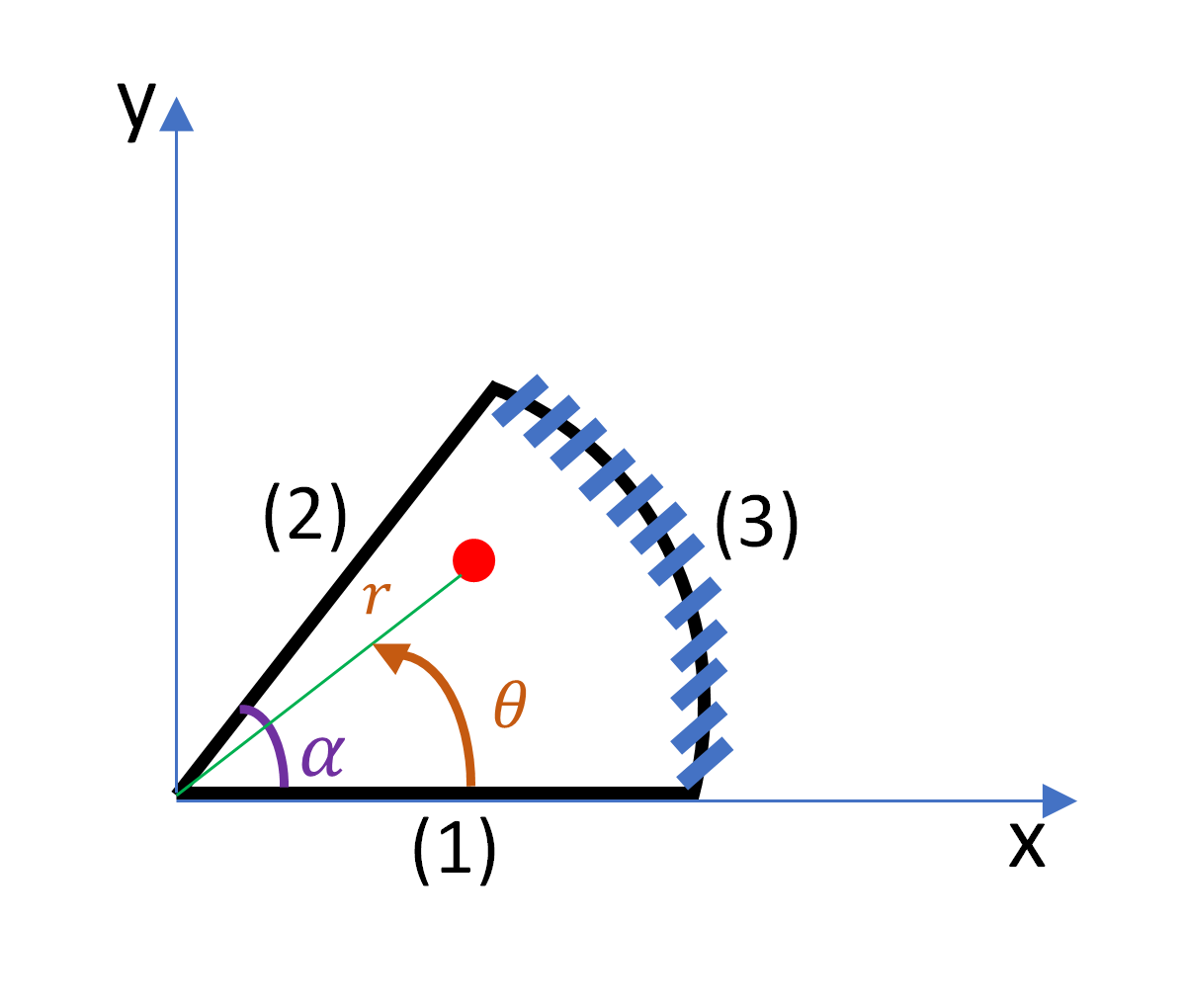}
    \caption{A schematic view of the sector boundary and a field located inside. In this figure the bow of the sector (side $3$) is reflective while the other two sides are absorbing.}
    \label{Boundariessec}
\end{figure}

Now we consider the case where the field is located  in a shape which is like a sector of a circle with radius $R$ and the central angle $\alpha$, see  Fig. \ref{Boundariessec} for a schematic view.  In this figure the side 3 (the bow) is reflective while the other two sides are absorbing. 

As  shown in \cite{Assadullahi:2016gkk} the average crossing time ,$\langle \mathcal{N}\rangle$, satisfies 
\begin{equation}
\label{averageNpde}
    \nabla^2\left<\mathcal{N}\right>=-2 \, .
\end{equation}
Note that on the absorbing  boundaries of the sector $\left<\mathcal{N}\right>=0$, i.e the boundary conditions are Dirichlet, while on the reflective boundaries we have the Neumann boundary condition $\nabla \langle\mathcal{N} \rangle.\hat{n}=0$ where $\hat{n}$ is the orthogonal unit vector to the boundary.

To solve the Laplace equation \eqref{averageNpde} one can follow the same procedure as in electromagnetic theory  and use the Green function method \cite{Jackson}. In other words one can treat the right hand side of 
Eq. \eqref{averageNpde} as the electrical charge density and $\left<\mathcal{N}\right>$ as the potential which satisfies the appropriate boundary conditions. Hence we solve the following equation in polar coordinate $(r, \theta)$: 
\begin{equation}\label{Laplace1}
    \nabla^{2} G(r,\theta)=\frac{\delta(r-r') \delta(\theta-\theta')}{r} \, .
\end{equation}

In what follows we study the setup with two different boundary conditions in details.  
 First, we consider the case where 
 both sides of the sector are absorbing boundaries and second, the case 
in which one of the sides is absorbing while the other one is reflective.  Note that in both cases the bow of the sector (side 3) is a reflective boundary. This study can be generalized to the case where the bow is an absorbing boundary as well 
but  for the sake of brevity we do not present the analysis for this case.

\vspace{1cm}
\begin{itemize}
  \item \textbf{Absorbing boundaries}
\end{itemize}

Suppose both sides of the sector (sides at $\theta=0$ and $\theta=\alpha$) are absorbing boundaries. 
To find the solution, we first calculate the Green function associated to the Laplace equation \eqref{Laplace1}, which for  the two regions $r<r' $ and $r>r'$, are given by 
\begin{equation}\label{G1-sector}
    G_{(r<r')}=\sum_{k=0}^{\infty} A_k r^{\frac{k \pi}{\alpha}}\sin{\big(\frac{k \pi}{\alpha}\theta\big)} \, ,
\end{equation}
and
\begin{equation}\label{G2-sector}
    G_{(r>r')}=\sum_{k=0}^{\infty} \left(\tilde{A}_k r^{\frac{k \pi}{\alpha}}+ \tilde{B}_k r^{-\frac{k \pi}{\alpha}} \right)\sin{\big(\frac{k \pi}{\alpha}\theta\big)} \, .
\end{equation}

After imposing the appropriate boundary conditions on the above two functions, 
the coefficients of the Green function \eqref{G1-sector} and \eqref{G2-sector} read
as 
\begin{equation}
    A_{k}=-\frac{1}{k \pi}\left[\left(\frac{r'}{R}\right)^{\frac{2 k \pi}{\alpha}}+1\right]r'^{-\frac{k \pi}{\alpha}}\sin{\big(\frac{k \pi}{\alpha}\theta'\big)}\,\,,
\end{equation}
and
\begin{equation}
    \tilde{A}_{k}=  \tilde{B}_{k}R^{-2(\frac{k\pi}{\alpha})}=-\frac{1}{k \pi}\left(\frac{r'}{R^2}\right)^{\frac{k\pi}{\alpha}}\sin{\big(\frac{k \pi}{\alpha}\theta'\big)}\,.
\end{equation}
Having the Green function at hand,  the solution of Eq. \eqref{averageNpde} can be written  as 
\begin{equation}
    \left<\mathcal{N}\right>=-2\int_0^{r}\int^\alpha_0G_{(r'<r)}r'dr'd\theta' -2 \int_{r}^R\int^\alpha_0G_{(r'>r)}r' dr' d\theta'.
\end{equation}
Correspondingly,  the average time it takes for the field to be absorbed by  either of the boundaries reads as
\begin{equation}
\label{avN-Sector-abs}
    \langle \mathcal{N} \rangle=
    -4 \alpha ^2\sum_{k=1}^\infty \Big[\frac{ k \pi r^2-2 \alpha  R^2 \left(\frac{r}{R}\right)^{\frac{k \pi}{\alpha }}}{\pi ^2 k^2 \left(k^2 \pi ^2 -4 \alpha ^2\right)}\Big]\big(\cos (k \pi  )-1\big) \sin \big(\frac{k \pi}{\alpha }\theta\big)\,,
\end{equation}
in which $(r, \theta)$ represent the initial position of the field in polar coordinate. 
From the above expression, we see that only the odd modes contribute, $k= 2n+1$
with $n=0, 1, ... \, $. 

\begin{figure}
    \centering
    \begin{subfigure}[t]{0.45\textwidth}
        \centering
        \includegraphics[width=\linewidth]{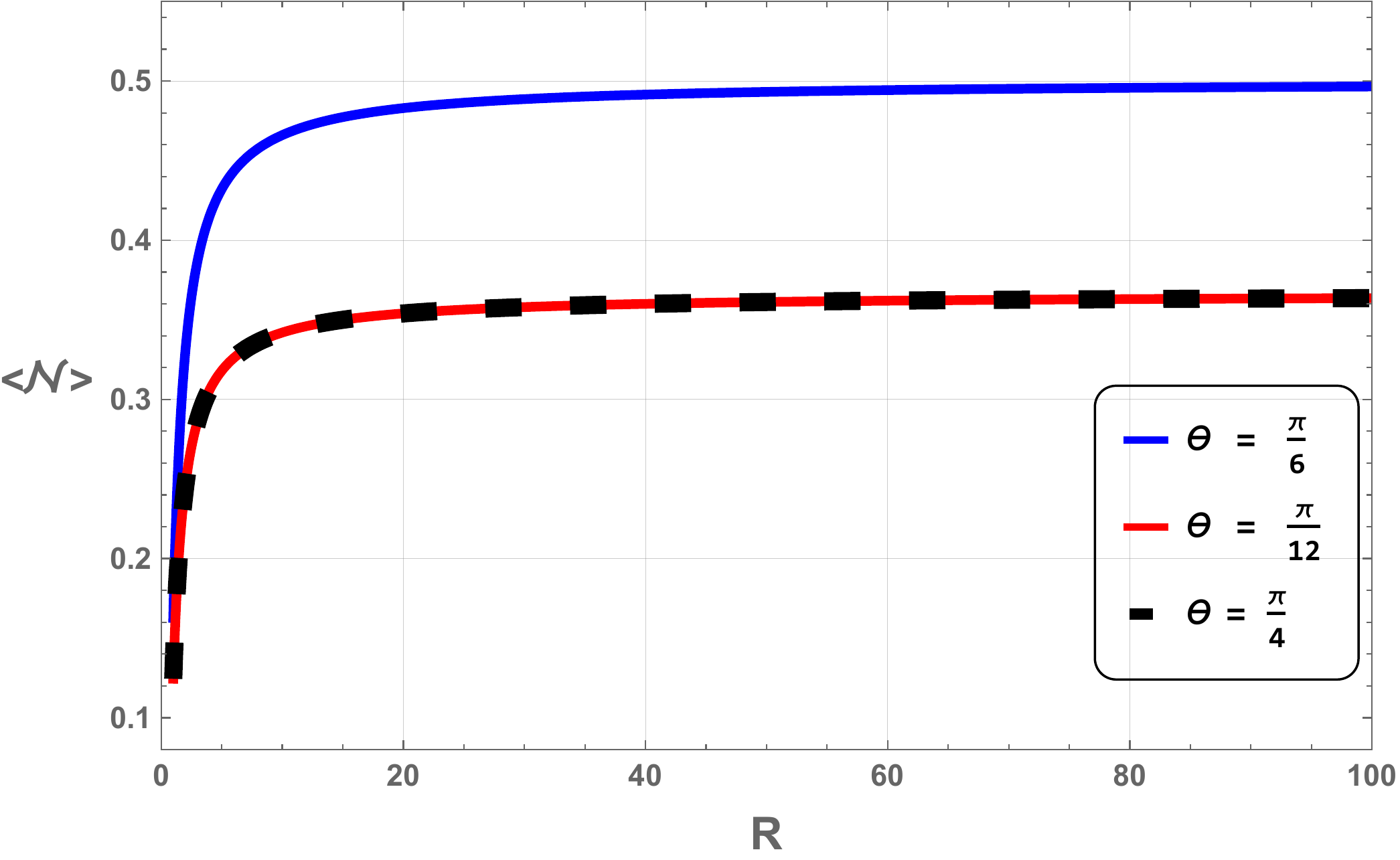}
    \end{subfigure}
    \hfill
    \begin{subfigure}[t]{0.49\textwidth}
        \centering
        \includegraphics[width=\linewidth]{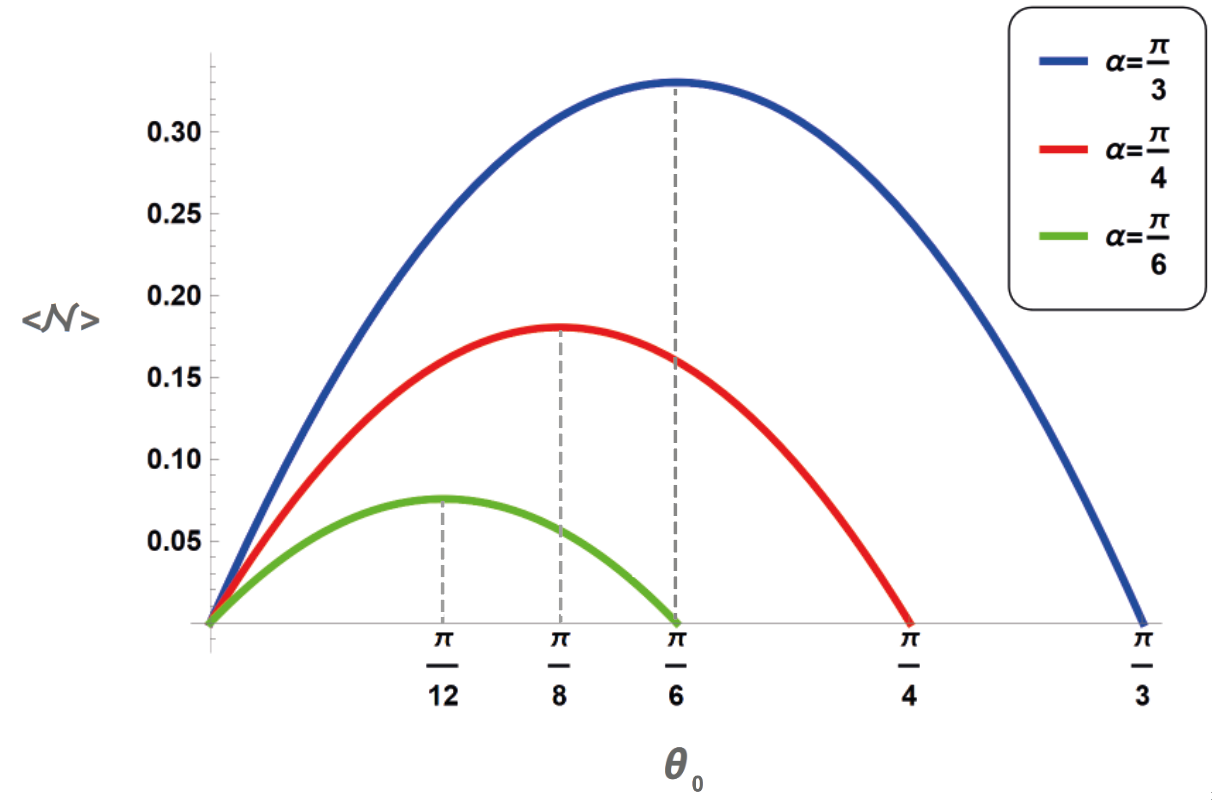}
    \end{subfigure}
    \caption{Left: $\langle\mathcal{N}\rangle$ vs. $R$ with $\alpha = \frac{\pi}{3}$
    and at a fixed initial radial position $r$ for various values of angular position $\theta$ in the sector with both sides being absorbing.   
    We see that the results for $\langle\mathcal{N}\rangle$ coincide 
for the initial positions $\theta= \frac{\pi}{12}$  and $\theta= \frac{\pi}{4}$  which are below the values for the case of the bisector $\theta= \frac{\pi}{6}$. 
Right: $\langle\mathcal{N}\rangle$ vs. $\theta$ with fixed values of $R$ and $r$
for various values of $\alpha < \frac{\pi}{2}$. We see that $\langle\mathcal{N}\rangle$ is finite with its maximum occurring at the position of bisector $\theta= \frac{\alpha}{2}$.   }
    \label{N-R-sector-abs}
\end{figure}

\begin{figure}
    \centering
    \begin{subfigure}[t]{0.48\textwidth}
        \centering
        \includegraphics[width=\linewidth]{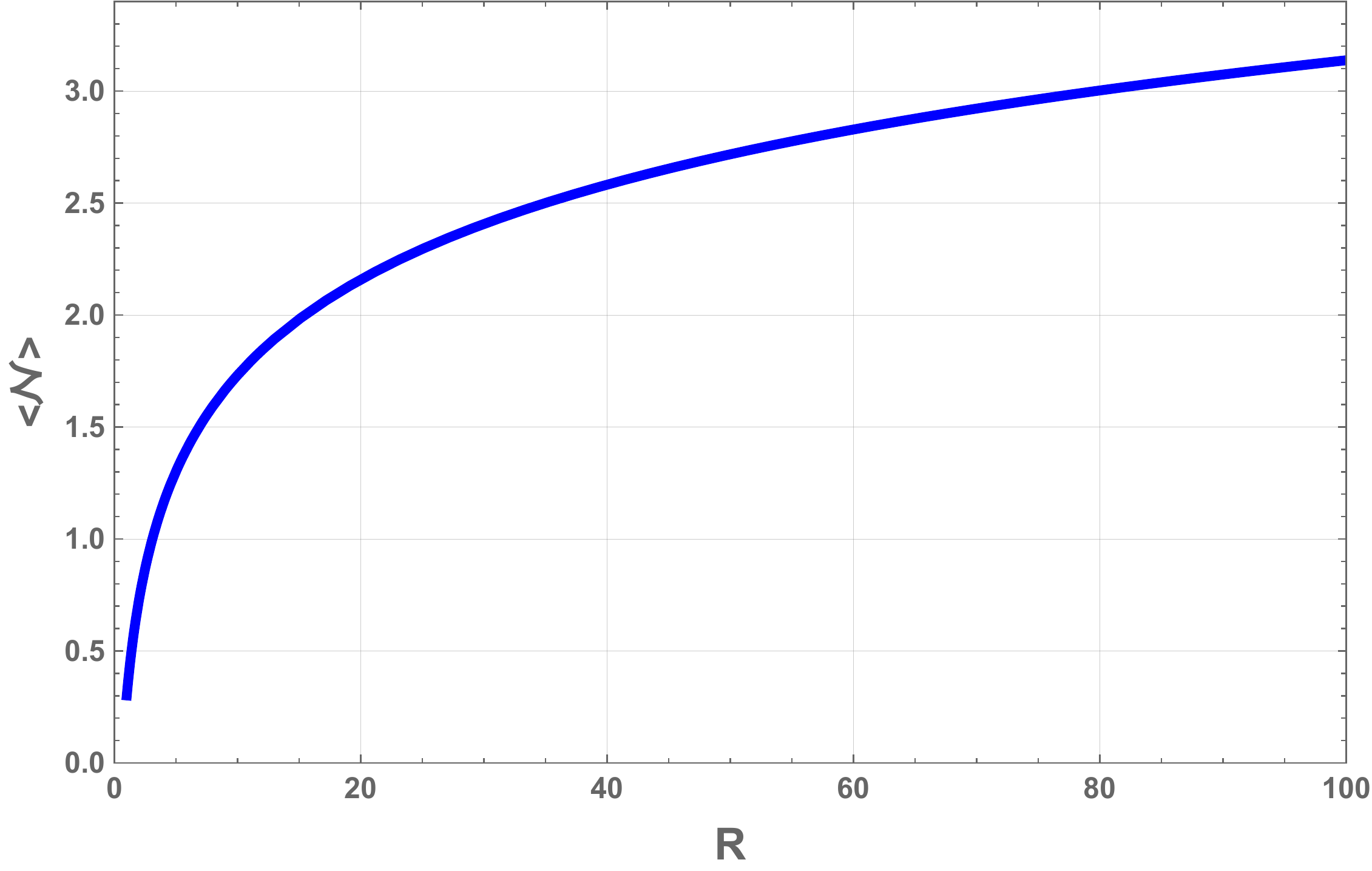}
    \end{subfigure}
    \hfill
    \begin{subfigure}[t]{0.48\textwidth}
        \centering
        \includegraphics[width=\linewidth]{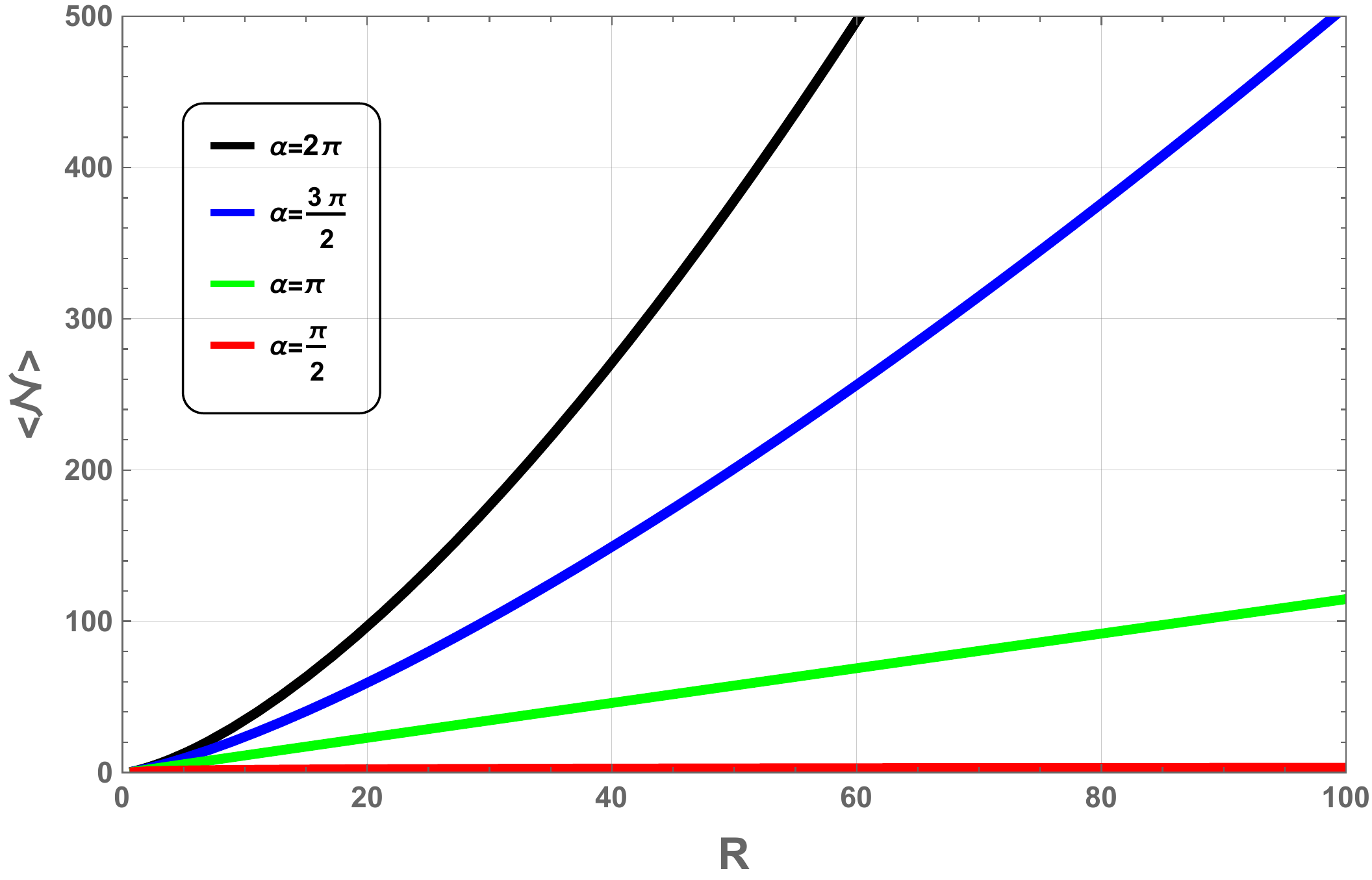}
    \end{subfigure}
    \caption{The behaviour of $\langle\mathcal{N}\rangle$ versus the radius of the sector, $R$, for sector with both sides being absorbing.
    In the left panel we consider $\theta = \frac{\pi}{4}$, $r= 1$ and  $\alpha =\frac{\pi}{2}$ while the right panel is for different angles, $\alpha \geq \frac{\pi}{2}$ ,with the same initial conditions for $r$ and $\theta$.  In all cases $\langle\mathcal{N}\rangle$ diverges, with the logarithmic divergence in the left panel and more rapid divergences in the right panel. }
    \label{N-R-sectors-abs}
\end{figure}

There are a number of interesting properties which can be deduced from 
Eq. (\ref{avN-Sector-abs}). First, since only the odd modes contribute in the above sum, one can show that the two initial conditions $\theta<\alpha$ and $\alpha - \theta$ with equal radial position $r$ yield the same value of $\langle \cN \rangle$. This is expected since these two points of initial conditions are at equal distances from their closest side of the sector.  Second, the maximum value of $\langle \cN \rangle$ takes place at the position of bisector, $\theta = \frac{\alpha}{2}$. 
Third, consider the limit that  $R\longrightarrow\infty$ so the sector becomes open.
In this limit, one can show that  $\langle \cN \rangle$ converges for the configuration 
in which $\alpha < \frac{\pi}{2}$ while for $\alpha \geq \frac{\pi}{2}$, $\langle \cN \rangle$ diverges with a logarithmic divergence for the special case of 
$\alpha = \frac{\pi}{2}$. All these conclusions can be seen in 
Figs. \ref {N-R-sector-abs} and \ref{N-R-sectors-abs}. 
In the left panel of Fig. \ref {N-R-sector-abs}  we have set $\alpha = \frac{\pi}{3}$ and looked at the behaviour of $\langle \cN \rangle$ for different values of the initial angular position  $\theta$ but with the same radial position $r$. 
The two initial conditions $\theta= \frac{\pi}{4}$ and $\theta= \frac{\pi}{12}$ yield the same results for $\langle \cN \rangle$ which are below the result for the  case of bisector with $\theta= \frac{\pi}{6}$. In the right panel of Fig. \ref {N-R-sector-abs}, we have looked at $\langle \cN \rangle$ as a function of $\theta$ for  fixed values of $R$ and $r$ and for 
different values of $\alpha< \frac{\pi}{2}$. We see that the maximum of $\langle \cN \rangle$ occurs at the position of bisector $\theta = \frac{\alpha}{2}$ with finite values of $\langle \cN \rangle$. In Fig.  \ref{N-R-sectors-abs}, $\langle \cN \rangle$ 
vs. $R$ is plotted for various cases of $\alpha \geq \frac{\pi}{2}$. The logarithmic 
divergence of $\langle \cN \rangle$ can be seen in the left panel for $\alpha= \frac{\pi}{2}$ while more rapid divergences can be seen in the right panel for $\alpha > \frac{\pi}{2}$. 

We can also calculate the crossing probability associated to each boundary. Using the same procedure as \cite{Assadullahi:2016gkk} one can show that the first crossing probability $p_i$ for $i=1, 2$ satisfies $\nabla^2p_i=0$. One can easily check that the following formula satisfies the corresponding Laplace equation for the  first hitting probability with appropriate boundary conditions, 
\begin{equation}
    p_1=-\frac{\theta}{\alpha}+1\,,
\end{equation}
while for the boundary 2 it is $p_2=1-p_1$. As one expects, the probability in this case is only a function of the angular position of the  field and is independent of 
its radial distance, $r$. An interesting case is the one in which $\alpha=2\pi$ where the two sides coincide each other. In this case the crossing probability of the side on $\theta'=0$ from the above formula is given by
\begin{equation}
	p_{1}=-\frac{\theta}{2\pi}+1 \,.
\end{equation}

We can also calculate the conditional average time $\langle \mathcal{N}_{i} \rangle$ to first cross  each of the absorbing boundaries $i=1, 2$. The conditional average time $\langle \mathcal{N}_{i} \rangle$ has the following interpretation.  Suppose that we have chosen all realizations in which the field has crossed the $i$-th boundary. 
Then the average time it takes for the field  to cross the $i$-th boundary is given 
by the ratio $\left<\mathcal{N}_i\right>/p_i$.  Following the same approach used in \cite{Assadullahi:2016gkk}, one can show  that $\langle \mathcal{N}_{i} \rangle$  obeys the following equation
\begin{equation}
    \nabla^2 \langle \mathcal{N}_{i} \rangle= -2 p_i \, .
\end{equation}
Thus, having $p_1$ and $p_2$ at hand, one can calculate $\langle \mathcal{N}_{i} \rangle$.

\begin{itemize}
  \item \textbf{Sector with mixed boundary conditions}
\end{itemize}

Now let us consider the case in which the boundary  at $\theta=0$ is reflective while the side at $\theta=\alpha$ is absorbing. As in previous case the third boundary, the bow   at $r=R$, is still reflective. We can define the surface of end of inflation (or end of USR) to be the absorbing boundary and calculate the power spectrum. For this purpose  we need to calculate $
\left<\mathcal{N}\right>$ and $ \left<\mathcal{N}^2\right>$.
The former is obtained by solving the Laplace Eq.  \eqref{averageNpde} while the latter satisfies the following equation \cite{Assadullahi:2016gkk}: 
\begin{equation}
    \nabla^2\left<\mathcal{N}^2\right>=-4\left<\mathcal{N}\right> \, .
\end{equation}

Using the method of Green function  one can see that the Green function in the sector with the current boundary conditions is given as follows:
\begin{equation}
    G_{(r<r')}=\sum_{k=0}^{\infty} A_k r^{\frac{ (2 k+1) \pi}{2 \alpha }} \cos \Big(\frac{(2 k+1)\pi }{2 \alpha } \theta\Big) \, ,
\end{equation}
and
\begin{equation}
    G_{(r>r')}=\sum_{k=0}^{\infty} \left(B_k r^{\frac{(2 k+1)\pi}{2 \alpha }}+C_k r^{-\frac{  (2 k+1)\pi}{2 \alpha }}\right)\cos \Big(\frac{ (2 k+1)\pi   }{2 \alpha }\theta\Big) \, ,
\end{equation}
where
\begin{equation}
    A_{k} = -\frac{2 \cos \Big(\frac{\pi  \theta _0 (2 k+1)}{2 \alpha }\Big) \Big(r'^{-\frac{\pi  (2 k+1)}{2 \alpha }}+\left(\frac{r'}{R^2}\right){}^{\frac{\pi  (2 k+1)}{2 \alpha }}\Big)}{\pi  (2 k+1)} \, ,
\end{equation}
and 
\begin{equation}
    B_{k} = C_{k} R^{\frac{ - (2 k+1)\pi}{ a}} =-\frac{2 \cos \Big(\frac{ (2 k+1)\pi  \theta _0}{2 \alpha }\Big) \left(\frac{r'}{R^2}\right){}^{\frac{  (2 k+1)\pi}{2 \alpha }}}{\pi  (2 k+1)} \, .
\end{equation}
Having calculated the Green function we obtain $\langle\mathcal{N}\rangle$ as follows: 
\begin{equation}
    \langle\mathcal{N}\rangle  =\sum_{k=0}^{\infty}  \frac{32 \alpha^2(-1)^k   \left(\pi  (2 k+1) r^2-4 \alpha  R^2 \left(\frac{r}{R}\right){}^{\frac{2 \pi  k+\pi }{2 \alpha }}\right)}{(2 \pi  k+\pi )^2 \left((2 \pi  k+\pi )^2-16 \alpha ^2\right)}\cos \Big(\frac{ (2 k+1)}{2 \alpha }\pi  \theta \Big).
\end{equation}

\begin{figure}[t]
    \centering
   \includegraphics[scale=0.55]{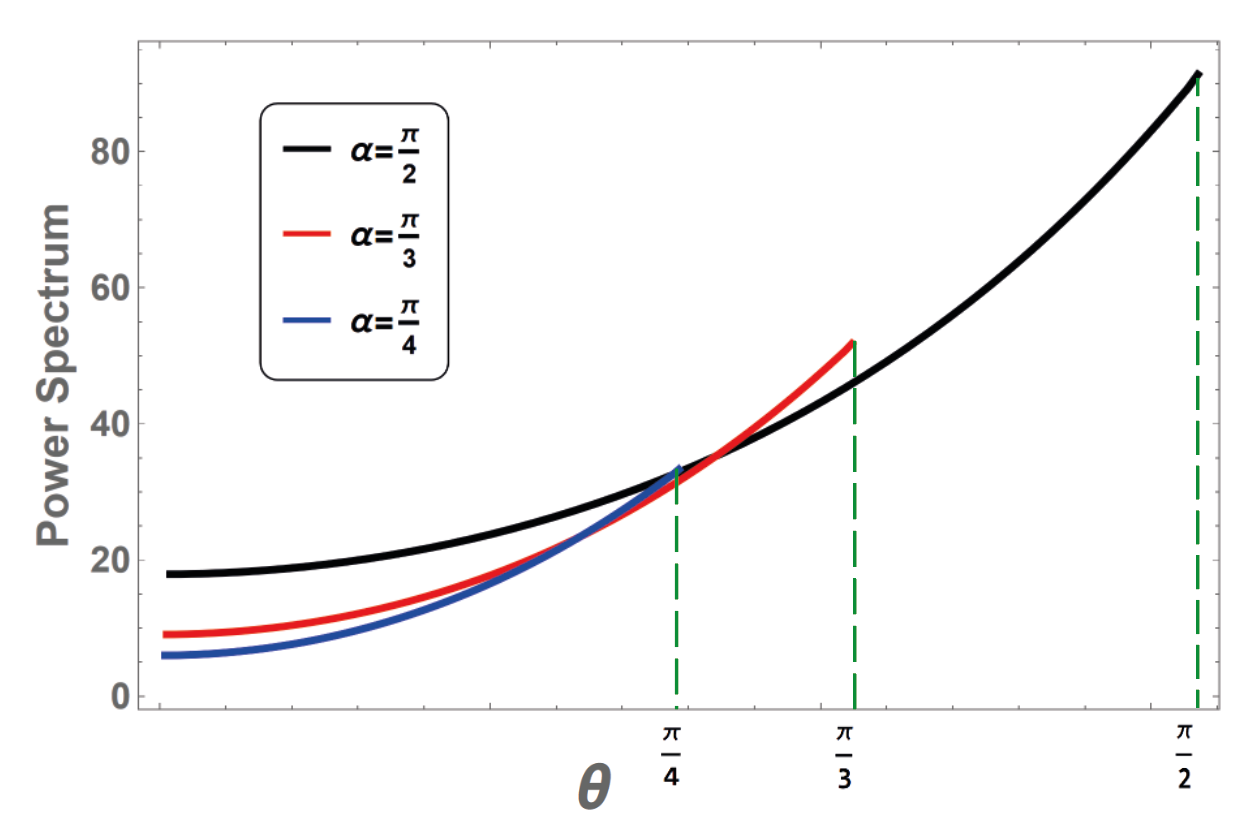}
    \caption{The power spectrum for various sectors with mixed absorbing and reflective edges  as a function of  $\theta$ . Here $r=6$ and $R=10$ in units of 
    $\frac{H}{2 \pi}.$  }
    \label{powerspectrum-sector}
\end{figure}
Having  $\left<\mathcal{N}\right>$ and the Green function at hand one can proceed to calculate $\left<\mathcal{N}^2\right>$ as follows
\begin{equation}
    \left<\mathcal{N}^2\right>=-4\int_0^\alpha\int_0^{r}\left<\mathcal{N}\right>G_{(r'<r)}r'dr'd\theta'-4\int_0^\alpha\int_{r}^{R}\left<\mathcal{N}\right>G_{(r'>r)}r'dr'd\theta' \, .
\end{equation}
For simplicity we avoid  presenting the explicit expression of $\langle\mathcal{N}^2\rangle$ here. 

Using the stochastic $\delta N$ formalism we calculate the power spectrum. 
 As the initial conditions have two degrees of freedom there is ambiguity about the direction of taking the derivative to calculate the  power spectrum. We take the derivative along the angular direction which is  orthogonal to the surface of end of inflation (i.e. the absorbing boundary), yielding, 
\begin{equation}
    \mathcal{P}_\calR=\frac{d\langle\delta\mathcal{N}^2\rangle}{d\left<\mathcal{N}\right>}=\frac{d\langle\delta\mathcal{N}^2\rangle}{d\theta}\frac{d\theta}{d\left<\mathcal{N}\right>} \, .
\end{equation}
The plot of the power spectrum can be found in Fig. \ref{powerspectrum-sector} which shows its behaviour versus the initial angular position of the field.

\begin{figure}[t!]
 \vspace{1cm}
    \centering
    \begin{subfigure}[t]{0.45\textwidth}
        \centering
        \includegraphics[width=\linewidth]{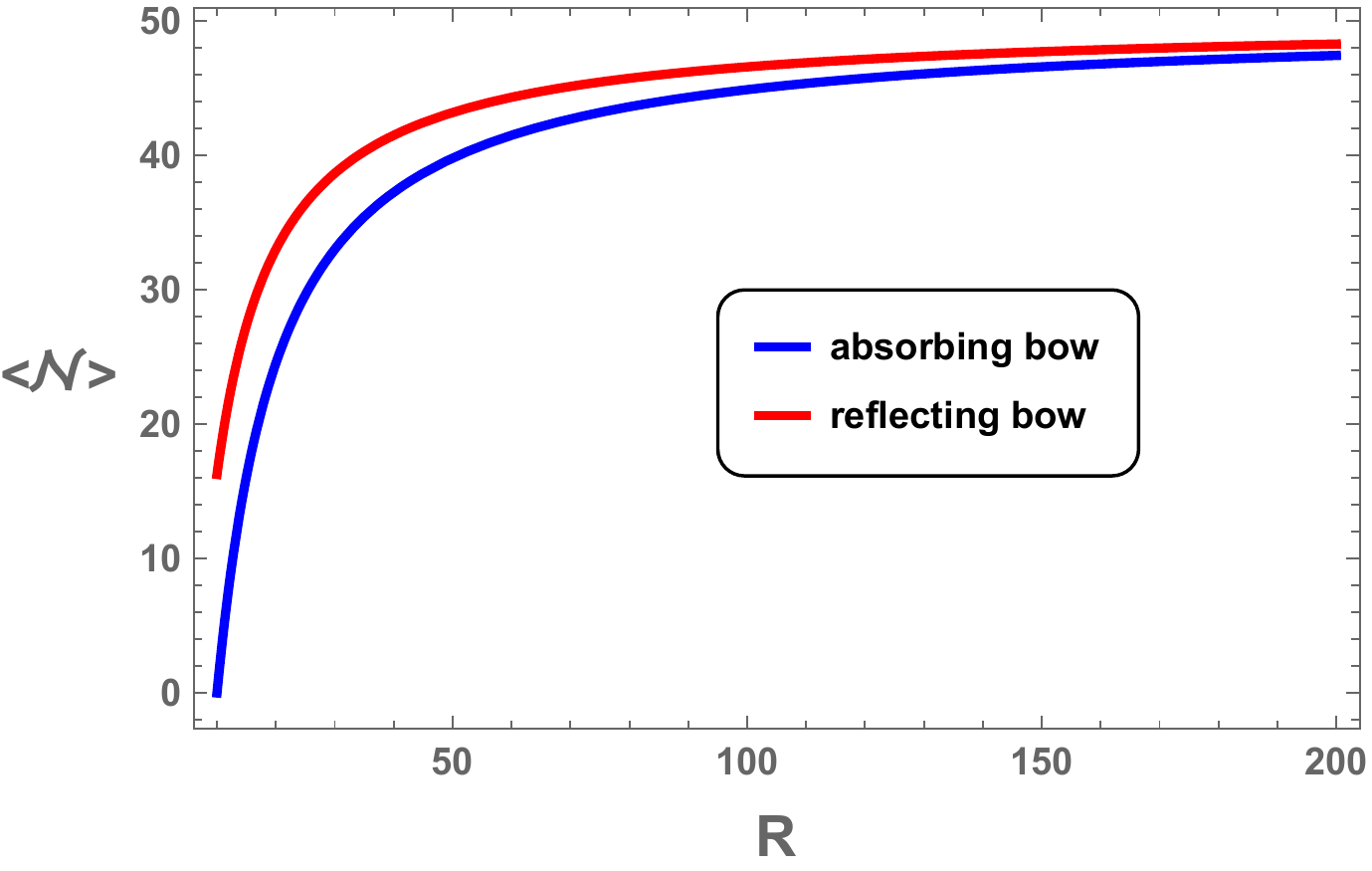}
    \end{subfigure}
    \hfill
    \begin{subfigure}[t]{0.45\textwidth}
        \centering
        \includegraphics[width=\linewidth]{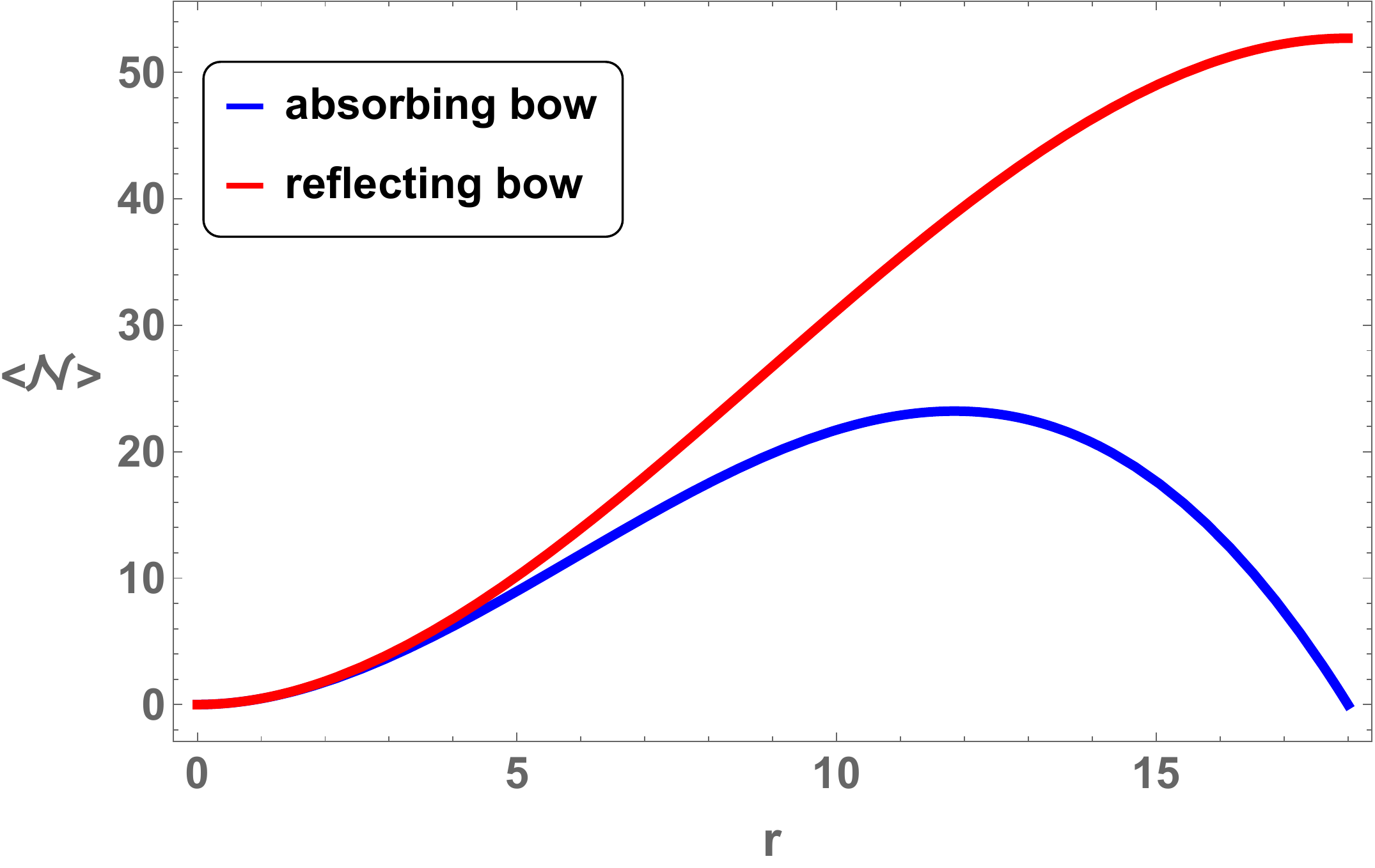}
    \end{subfigure}
    \caption{ Left: the behaviour of $\left<\mathcal{N}\right>$ versus $R$ when the bow of the sector is either reflective or absorbing
    for the fixed initial position $r=10$. Right panel: The behaviour of  $\left<\mathcal{N}\right>$ versus $r$ for a fixed value of $R=18$ when the bow of the sector is either reflective or absorbing.  In both panels we consider $\alpha=\frac{\pi}{3}$ and $\theta=\frac{\pi}{6}$.}
    \label{bowcompare}
\end{figure}

In the analysis so far we have assumed that the bow of the sector is reflective while the other two radial sides can be either fully absorbing or one being absorbing and the other one being reflective. For comparison, it would be interesting to consider 
the case where the bow is absorbing as well. We skip the detail analysis for brevity but in Fig. \ref{bowcompare} we have compared the corresponding results
when the bow is absorbing or reflective. In the left panel we see that 
$\langle \cN \rangle$ decreases if one switches the bow from being reflective to 
becoming absorbing.  This is easy to understand because 
the field hits the bow for time to time and  when the bow is absorbing the duration of inflation becomes shorter. However, as $R$ increases (the bow is far away) the difference between the two values of $\langle \cN \rangle$ becomes less significant. 
In the right panel, we have fixed the value of $R$ but have looked at $\langle \cN \rangle$ as a function of the initial position of the field $r$. As in the left panel, for a given value of $r$ the result for $\langle \cN \rangle $ in absorbing boundary is smaller than its value in the  case of reflective boundary. However, we also see a non-trivial trend for $\langle \cN \rangle$ as it develops a maximum at some intermediate  values of the initial position $r$ for the case where all three sides are absorbing.  This is easy to understand. If the field is too close to the bow, then it hits the absorbing bow quickly and inflation ends. On the other hand, if it is too far from the bow, it will hit the radial absorbing boundaries and inflation ends quickly as well. The maximum amount of inflation takes place when the field is initially somewhere in between when it takes a long time to hit either of the three absorbing boundaries.  

\section{Primordial Black Holes Formation }
\label{PBH}

There has been a revival of interest in PBHs in recent years \cite{Sasaki:2018dmp, Carr:2020gox, Green:2020jor, Byrnes:2021jka} after the discovery of gravitational waves from merging  black holes with mass at the order of
tens of solar mass  in the LIGO/Virgo observations  
\cite{LIGOScientific:2016aoc, LIGOScientific:2016sjg}. As black holes with this range of mass may not form from the known astrophysical processes, it is argued that these objects may indeed be PBHs. On the other hand, PBHs are extensively studied  as   candidate for dark matter \cite{Carr:2016drx}.   In order for PBHs to survive the Hawking radiation to furnish all or a fraction of observed dark matter, they should be heavier than about  $10^{15} {\mathrm {gr} }$. While light PBHs ($10^9 {\mathrm {gr} } < M < 10^{16} {\mathrm {gr} }$) are constrained from the effects of  Hawking evaporation on big bang nucleosynthesis and extra galactic background photons, the heavier PBHs with 
$10^{16} {\mathrm {gr} } < M < 10^{50} {\mathrm {gr} }$ are constrained by various gravitational effects such as microlensing  \cite{Sasaki:2018dmp, Carr:2020gox}. 
 Typically, PBHs are formed during radiation dominated era from the collapse of an overdense region. More specifically, a PBH will form when the cosmic density contrast $\delta =\frac{\delta \rho}{\rho}$ associated to a scale which re-enters the horizon  exceeds a critical value $\delta_c\sim  c_s^2$ in which $c_s$ is the sound speed of scalar perturbations during radiation dominated era \cite{CarrPBH:1975 Apj}. The mass of the formed PBH is at the order of the horizon mass  when the corresponding scale enters the horizon, $M \sim t/G$ in which $G$ is the Newton constant. 
 
Usually, the abundance of PBHs is given by the parameter $\beta$
which is the fraction of the energy density in PBHs to the total energy density at the time of formation.  Considering PBHs of masses between $M$ and $M + d \ln{M}$ at the time of formation, one can define $\beta(M) d\ln{M}$ as the fraction of the
mass density of the universe comprised in such PBHs. The mass fraction of PBHs against the total dark matter density at the present time is given by
\begin{equation}
\begin{split}
    f_{PBH} (M) =\frac{\Omega_{PBH}}{\Omega_{DM}}= 2.7 \times 10^8 \Big(\frac{M_{PBH}}{\textup{M}_\odot}\Big)^{\frac{-1}{2}}\beta(M)\, ,
\end{split}
\end{equation}
in which $\textup{M}_\odot = 2 \times 10^{30} \mathrm {gr}$ is the solar mass. 
The observational constraints for the heavy PBHs in the mass range $10^{16} {\mathrm {gr} } < M < 10^{50} {\mathrm {gr} }$  yield $\beta < 10^{-11} $ to  $\beta < 10^{-5} $ while for 
light PBHs with $10^{9} {\mathrm {gr} } < M < 10^{16} {\mathrm {gr} }$ the  observational constraints yield $\beta < 10^{-24} $ to  $\beta < 10^{-17} $ 
\cite{Carr:2009, Carr:2017, Carr:2020}.

 On large CMB scales, the amplitude of curvature perturbations are small, 
$\calP_\calR \sim 10^{-9}$ \cite{Planck:2018vyg, Planck:2018jri}. On the other hand, on smaller scales when the perturbations re-enter the Hubble radius during radiation dominated era they can become large to overcome the pressure gradients and form PBHs. Therefore, it is interesting to study the inflationary fluctuations non-perturbatively. If the mean curvature perturbation in a given Hubble patch 
exceeds  a critical value $\calR_c\sim 1$ then PBHs can form. 

The mass fraction of the PBHs is determined by the probability that the mean value
of curvature perturbation inside a Hubble patch exceeds  $\calR_c$, which from the 
 Press-Schechter formalism is given as \cite{Carr:2017}
\begin{equation}\label{beta_difinition}
  \beta(M) \sim  \int_{\calR_c}^{\infty} P({\calR}) d\calR\, .
\end{equation}
For a Gaussian PDF, the above mass fraction is determined by the curvature perturbation power spectrum  $\mathcal{P}_{\calR}$ which controls the variance of PDF (note that in our notation $P({\calR})$ is the PDF while $\mathcal{P}_{\calR}$ is the power spectrum which should not be confused with each other). 

The USR setup has been employed extensively in recent years as a mechanism to generate PBHs during inflation, for an incomplete list of papers on this active direction see  \cite{Ivanov:1994pa, Garcia-Bellido:2017mdw, Biagetti:2018pjj, Franciolini:2018vbk, Motohashi:2017kbs, Germani:2017bcs, Ragavendra:2020sop, Ozsoy:2021pws, Hooshangi:2022lao, Cai:2022erk, Cai:2021zsp,  Pi:2022ysn}. This is because during the USR phase of inflation the curvature perturbation is not frozen and grows like $\calR \propto a(t)^3$ \cite{Namjoo:2012aa, Chen:2013aj, Martin:2012pe, Akhshik:2015rwa} in which $a(t) \propto e^{H t}$ is the scale factor during inflation. As a result, for a finite period of USR inflation the amplitude of $\calP_\calR$ can increase by about a factor 
$10^{7}$ compared to CMB scales. Of course, in order to have a successful inflationary mechanism, the USR phase should be terminated followed by a conventional period of slow-roll inflation. Furthermore, in order not to conflict with  the observations on CMB scales, it is also assumed that the USR phase is preceded by an earlier phase of slow-roll inflation as well.   We comment that recently it is argued that  the modes which leave the horizon during the USR phase may induce  large  one-loop corrections on CMB scale modes  
\cite{Kristiano:2022maq, Kristiano:2023scm}, see also \cite{Cheng:2021lif}. 
However, it is argued in 
\cite{Riotto:2023gpm, Riotto:2023hoz} that if one considers a mild transition the dangerous one-loop corrections are washed out, see also 
\cite{Choudhury:2023vuj, Choudhury:2023jlt, Choudhury:2023rks} on the effects of renormalizations in one-loop calculation. This question was  investigated explicitly in  \cite{Firouzjahi:2023aum}  and it was shown that the amplitude of the one-loop correction is controlled by the sharpness of the transition. The bottomline is that one can still use a period of USR phase to amplify the power spectrum for PBHs formation if the transition from the USR phase to the final slow-roll phase 
is smooth \cite{Riotto:2023gpm, Riotto:2023hoz}.

Since the potential is very flat during the USR phase, the quantum diffusion effects may become important and the system may not be perturbative \cite{Pattison:2019hef, Pattison:2021oen}. This is the regime  which was  studied in the previous sections of the current work where the classical drift term is neglected and the evolution of perturbations is entirely given by the quantum diffusion effects. Here we continue this analysis by calculating the PDF of curvature perturbations  during the quantum diffusion dominated period to obtain the mass fraction $\beta(M)$ and the fraction of PBHs in dark matter energy density. To simplify the analysis, we first consider the spherical symmetric configurations in various dimensions $n$ as studied in Section \ref{symmetric} and then investigate the case of 
sector boundary studied in section \ref{Section_Sector} as an example of the asymmetric boundary.

\subsection{PDF of $n$-Sphere} 

Starting with the characteristic function \eqref{Char1}, one can perform the inverse Fourier transform to obtain the PDF $P(\mathcal{N},r)$ as 
\begin{equation}
\label{Char2}
   P(\mathcal{N},r)=\frac{1}{2\pi}\int_{-\infty}^\infty e^{-i t \mathcal{N}} \chi_\mathcal{N}(t,r)dt\,.
\end{equation}
Following \cite{Ezquiaga:2019ftu, Pattison:2021oen} we expand  the characteristic function using the residue theorem as 
\begin{equation}\label{Char3}
    \chi_\mathcal{N}(t,r)=\sum_{k=0}^{\infty}\frac{a_{k}(r)}{\Lambda_{k}- it} + f(t,r) \, ,
\end{equation}
in which $f(t,r)$ is a regular function of $t$ and $\Lambda_{k}$'s are positive numbers independent of $r$. The PDF can then be given by
\begin{equation}\label{PDF-residue}
    P(\mathcal{N},r)=\sum_{k=0}^{\infty}{a_k(r)}e^{-\Lambda_{k}\mathcal{N}}.
\end{equation}
We note that for large values of $\mathcal{N}$ the first term in the above summation dominates over the next terms. This dominant term is determined  by $\Lambda_{0}$ and $a_{0}$, which are the lowest pole of the characteristic function and the corresponding residue, respectively. Contributions from higher terms of $\Lambda_{k}$ decay exponentially with a decay rate that can be given by solving the characteristic function from Eq. \eqref{Fourieradjioint} and finding the zeros of its inverse. The residues are also obtained as
\begin{equation}\label{an}
    a_{k}(r)= -i\Big[\frac{\partial}{\partial t} \chi_{\mathcal{N}}^{-1} \Big{|}_{t=-i\Lambda_{k}} \Big]^{-1}.
\end{equation}

Now, using the results from  Section \ref{symmetric} we can calculate the PDF in each case. Below we present the analysis for the cases $n=2$ and $n=3$ as examples.  

\begin{itemize}
  \item \textbf{$n=2$}
\end{itemize}
The characteristic function for the two-dimensional circular boundary with one reflective and one absorbing boundary is given by Eq. \eqref{charcircle} with coefficients $A$ and $B$ obtained in Eqs.  \eqref{coefref-A-circle} and \eqref{coefref-B-circle}. To continue, we study the setup where  $r_- \ll r_+$. In this limit  one can show that at leading order the first pole of the characteristic function is given by, 
\begin{equation}\label{pole_circle}
 \Lambda_0\simeq\frac{ 2.9}{ r_+^2}\, .
\end{equation}
We don't present the solution for $a_0(r)$ in this case as the corresponding formula 
has a  long and  complicated form. 
Correspondingly, the leading order PDF is given by  $P(\mathcal{N},r,\theta)=a_0(r,\theta)e^{-\Lambda_0 \mathcal{N}}$.

\begin{itemize}
  \item \textbf{$n=3$}
\end{itemize} 
The characteristic function in this case  with one reflective and one absorbing boundary is given by Eq. \eqref{char-nsphere} with the coefficients $A$ and $B$ obtained as in Eqs. \eqref{coefref-A-nsphere} and \eqref{coefref-B-nsphere} with $n=3$.  Similar to what we did in two-dimensional case we focus on the limit in which $r_- \ll r  \ll r_+$. In this limit we obtain 
\begin{equation}
\label{pole_sphere}
   \Lambda_0 \simeq \frac{\pi^2}{2 r_{+}^2} \,,
\end{equation}
while $a_0(r)$ is given by the following simple expression
\begin{equation}\label{a_sphere}
a_0(r)\simeq\frac{\pi  \sin \left(\frac{\pi  r}{r_+}\right)}{r r_+}.
\end{equation}
Correspondingly, up to the leading order in this limit  the analytic form of PDF from Eq. \eqref{PDF-residue} is obtained to be 
\begin{equation}
   P(\mathcal{N},r)\simeq \frac{\pi  \sin \left(\frac{\pi  r}{r_+}\right)}{r r_+}e^{-\frac{\pi^2}{2 r_{+}^2}\mathcal{N}} \, .
\end{equation}

\subsection{PDF of the Sector}
As an example of asymmetric boundaries, here we study the PBHs mass fraction for the example of sector boundary. For this purpose,   we first need to obtain the corresponding PDF. To this end we use a new method which is presented in detail in Appendix \ref{PDF-Sector}.  As a particular example, the PDF of a sector in which two boundaries (sides (1) and (3) of Fig. \ref{Boundariessec}) are considered reflective and just one boundary (side (2)) is absorbing, up to the leading order is given by 
\begin{equation}
\label{PDFsector}
    P(\mathcal{N})\simeq -\frac{2^{3-\frac{\pi }{2 \alpha }} r_{01}^{\frac{\pi }{2 \alpha }} \cos \left(\frac{\pi  \theta }{2 \alpha }\right) e^{-\frac{N r_{01}^2}{2 R^2}} J_{\frac{\pi }{2 \alpha }}\left(\frac{r r_{01}}{R}\right) \, _1F_2\left(\frac{\pi }{4 \alpha };1+\frac{\pi }{4 \alpha },1+\frac{\pi }{2 \alpha };-\frac{r_{01}^2}{4}\right)}{\alpha  R^2 \Gamma \left(1+\frac{\pi }{2 \alpha }\right) \left[-2 J_{\frac{\pi }{2 \alpha }}\left(r_{01}\right)+J_{\frac{\pi }{2 \alpha }-2}\left(r_{01}\right)+J_{2+\frac{\pi }{2 \alpha }}\left(r_{01}\right)\right] J_{\frac{\pi }{2 \alpha }}\left(r_{01}\right)}\,,
\end{equation}
where $r_{01}$ is the first root of $J'_{\frac{(2 m+1) \pi}{2 \alpha}}(r)$ with $m=0$. 

In the next subsection we calculate the PBHs mass fraction  for spherical boundaries with various values of $n$ as well as the case of sector and put constraints on the parameters of the models.

\subsection{  Mass Fraction }

In Section \ref{symmetric} we have calculated the average  number of e-folds  \eqref{N-sphereref} and  power spectrum \eqref{power_sphere} while in the previous subsection we have obtained  the PDF (for $n=2, 3$ and the example of sector). 
We are now in a position to study the PBHs formation and their mass fraction. Using the above mentioned equations and plugging them into the definition of the mass fraction Eq. \eqref{beta_difinition}, we obtain this parameter for a general value of $n$. 

\begin{figure}
    \centering
    \begin{subfigure}[t]{0.45\textwidth}
        \centering
        \includegraphics[width=\linewidth]{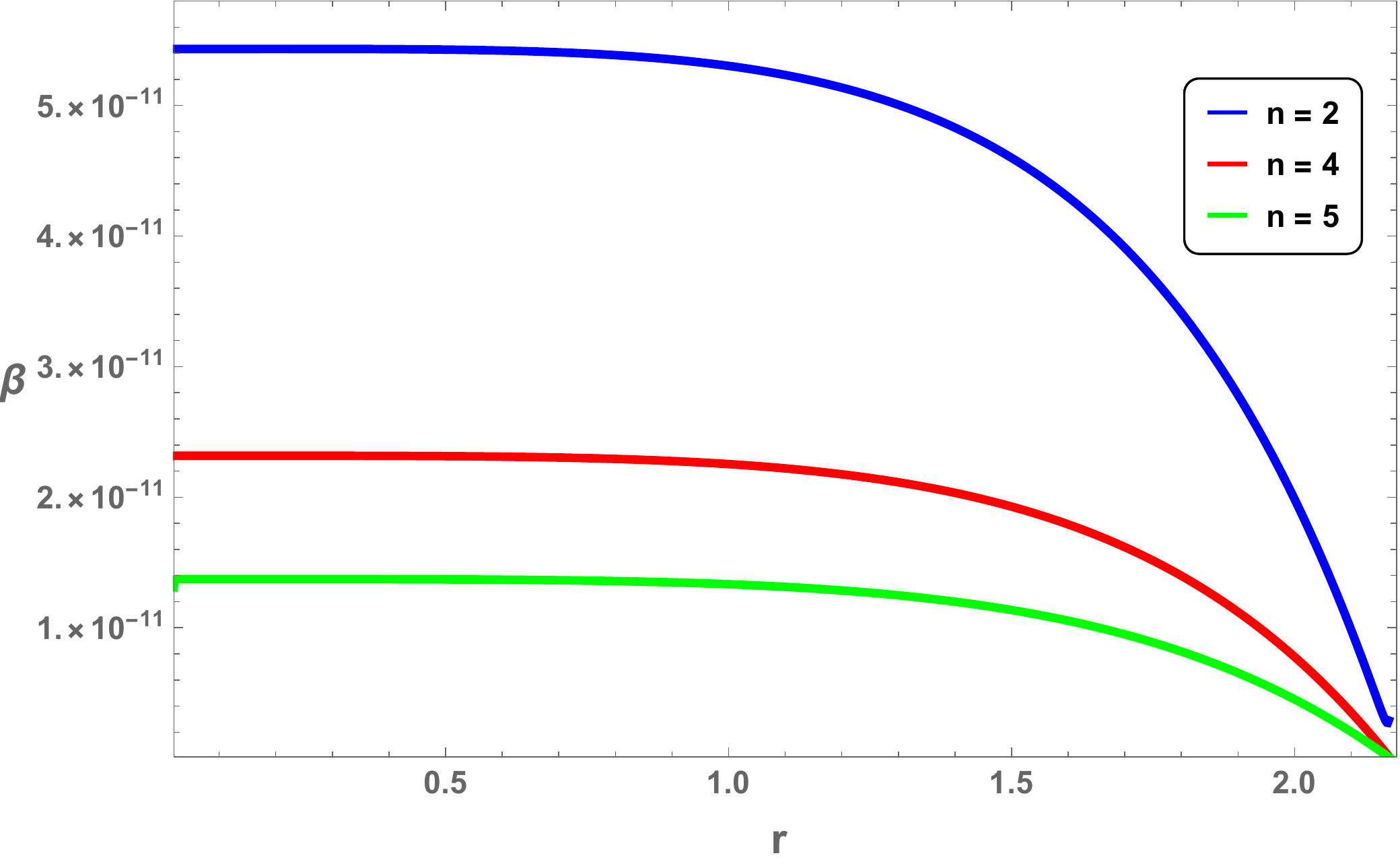}
    \end{subfigure}
    \hfill
    \begin{subfigure}[t]{0.45\textwidth}
        \centering
        \includegraphics[width=\linewidth]{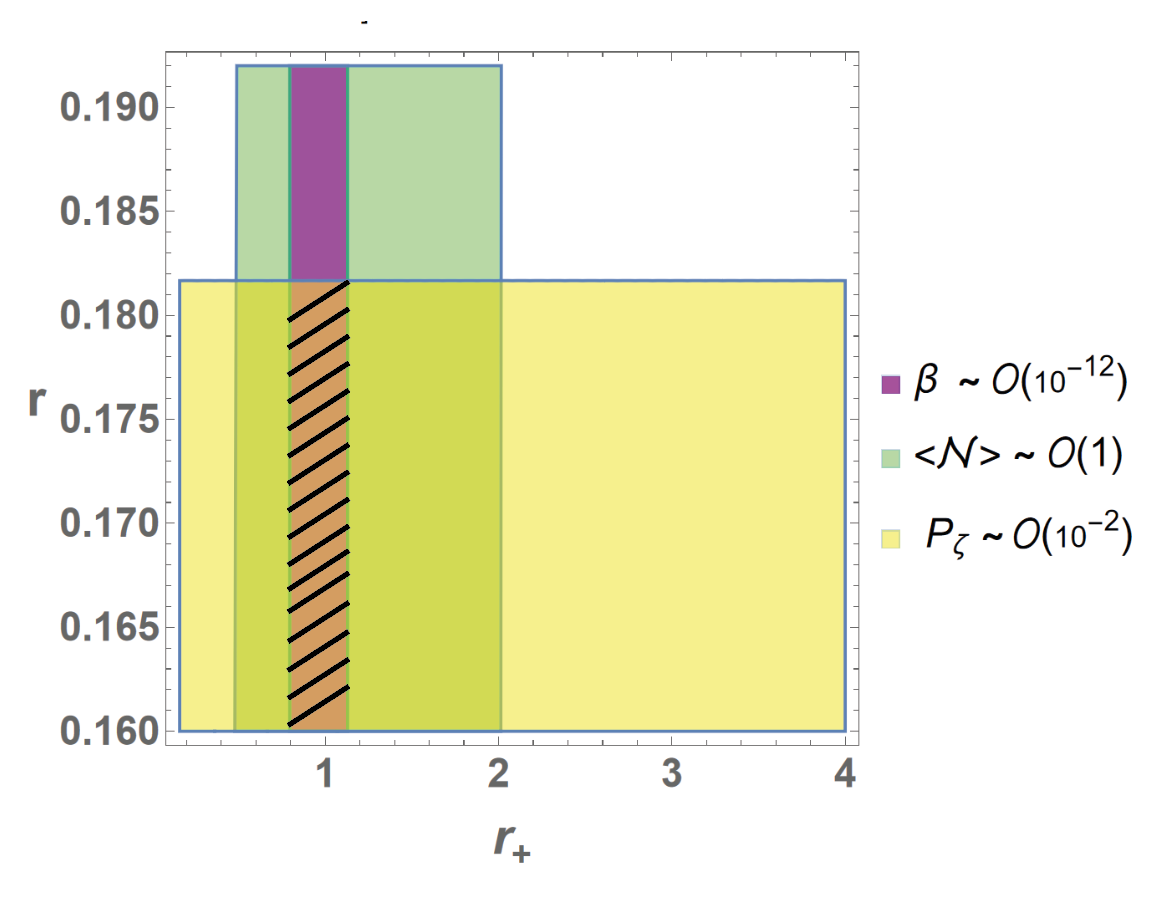}
    \end{subfigure}
    \caption{ Left: The behaviour of the mass fraction versus the initial position of the field in spherical boundaries for various values of $n$.  With equal initial conditions  $r_+ = 2.2$ and $r_- =0.03$ in units of $\frac{H}{2 \pi}$, increasing the dimensions of the field space yields  smaller values of $\beta$.   Right: the allowed ranges of 
 $\beta$, $\mathcal{P}_\calR $ and $\langle \cN \rangle$ for the case 
 $n=2$. The magenta colour is the region in which $\beta$ is in the acceptable range, the yellow colour area shows the region where  
 $\mathcal{P}_\calR \sim 10^{-2}$ and the green area shows region with acceptable values of $ \langle \cN \rangle$. The  area hatched with black lines represents the region of parameters in which all these three constraints are satisfied. } 
    \label{beta-nD}
\end{figure}

Let us start with the case of the spherical boundaries. 
As the analytic expression for $\beta(M)$ is complicated 
we do not present them here.
The behaviour of mass fraction versus the initial position of the field $r$ for various values of $n$ is presented  in the left panel of Fig. \ref{beta-nD}.  Considering fixed equal initial conditions for $r_+$ and $r_-$,  we see that  increasing the dimensions of the field space results in a smaller  value for $\beta$.  
To obtain the plot in the left panel of Fig. \ref{beta-nD}, we have imposed a number of requirements on our results for $\beta$, $\mathcal{P}_{\zeta}$ and $\langle\mathcal{N} \rangle$. Specifically, we demand  the USR phase to last
for a few e-folds, requiring $\langle\mathcal{N} \rangle \sim \mathcal{O}(1)$. This is because, as we discussed before, a USR phase is usually assumed to be sandwiched between two periods of slow-roll inflation. In addition, to have any chance for PBHs formation we actually require the power spectrum to be large enough, say $\mathcal{P}_{\calR} \sim \mathcal{O}(10^{-2})$ with   $\beta \lesssim 10^{-11}$. For this reason, we did not present the result for the case $n=3$ 
in the left panel of Fig. \ref{beta-nD} since the resultant values of $\beta$ did not match this requirement. For further illustration,  the right panel of Fig. \ref{beta-nD} shows the intersections of these three constraints  in the field space for the case $n=2$.  The yellow colour area  in this figure shows where the power spectrum satisfies  $\mathcal{P}_\calR \sim 10^{-2}$, the magenta colour region shows the acceptable range for $\beta$ and the green region corresponds to acceptable values of $\langle\mathcal{N}\rangle$ during the USR period. The hatched area shows the region for initial field value versus the absorbing boundary position 
$r_+$ for which all three constraints are satisfied.

\begin{figure}
 \vspace{.1cm}
    \centering
   \includegraphics[scale=0.55]{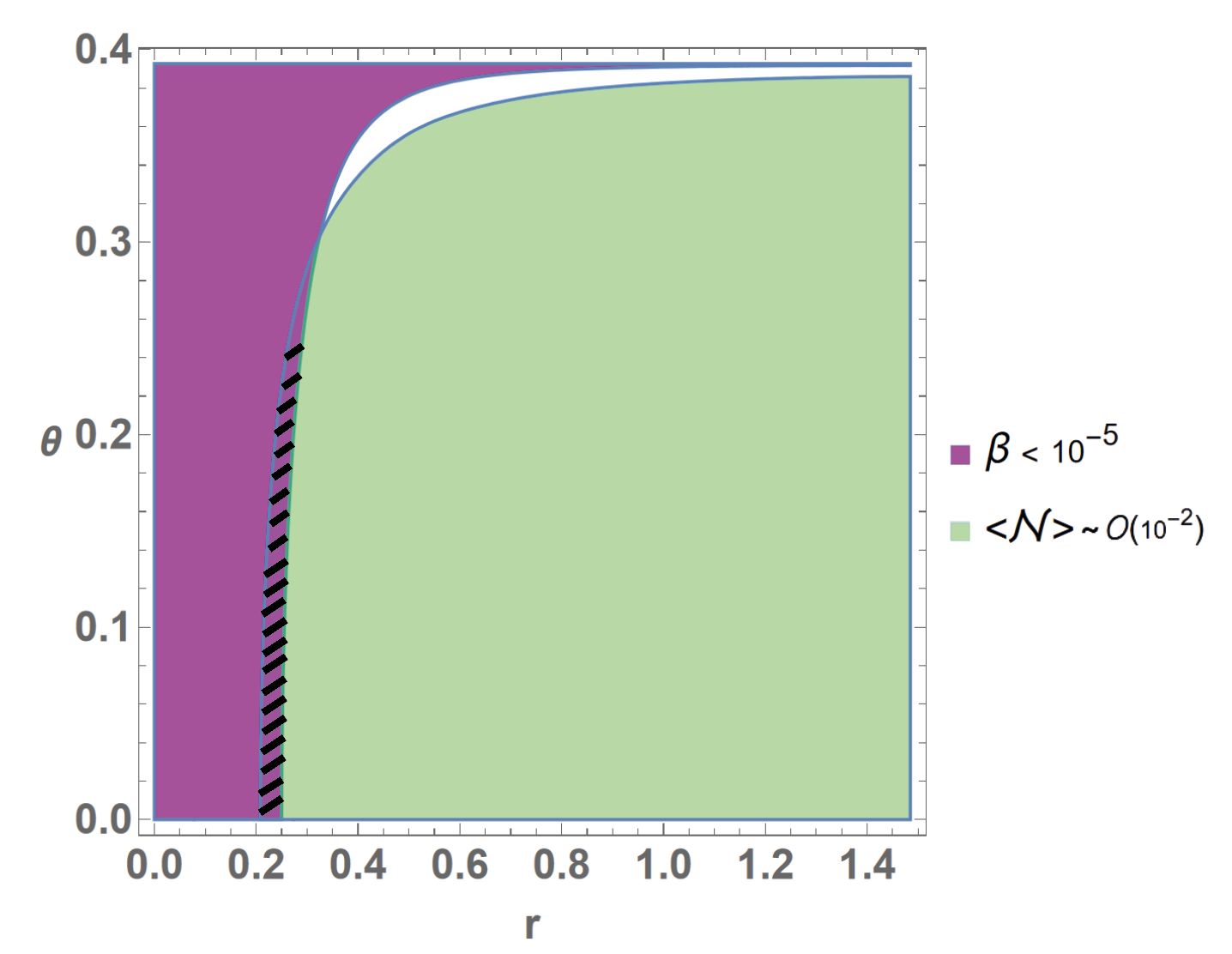}
    \caption{ The allowed ranges of 
    $\beta$ and $\langle \cN \rangle$ for a sector with mixed boundary conditions. Here $\alpha=\frac{\pi}{8}$ and $R=2$ in terms of $\frac{H}{2\pi}$. For these parameters $\mathcal{P}_{\mathcal{R}}\simeq O(0.01)$ in the whole region. The hatched area is where all these constraints are met.}
    \label{f_PBH-Sector}
\end{figure}

For a similar plot for the example of sector see Fig. \ref{f_PBH-Sector}. 

\begin{figure}[h]
 \vspace{-0.5cm}
    \centering
   \includegraphics[scale=0.55]{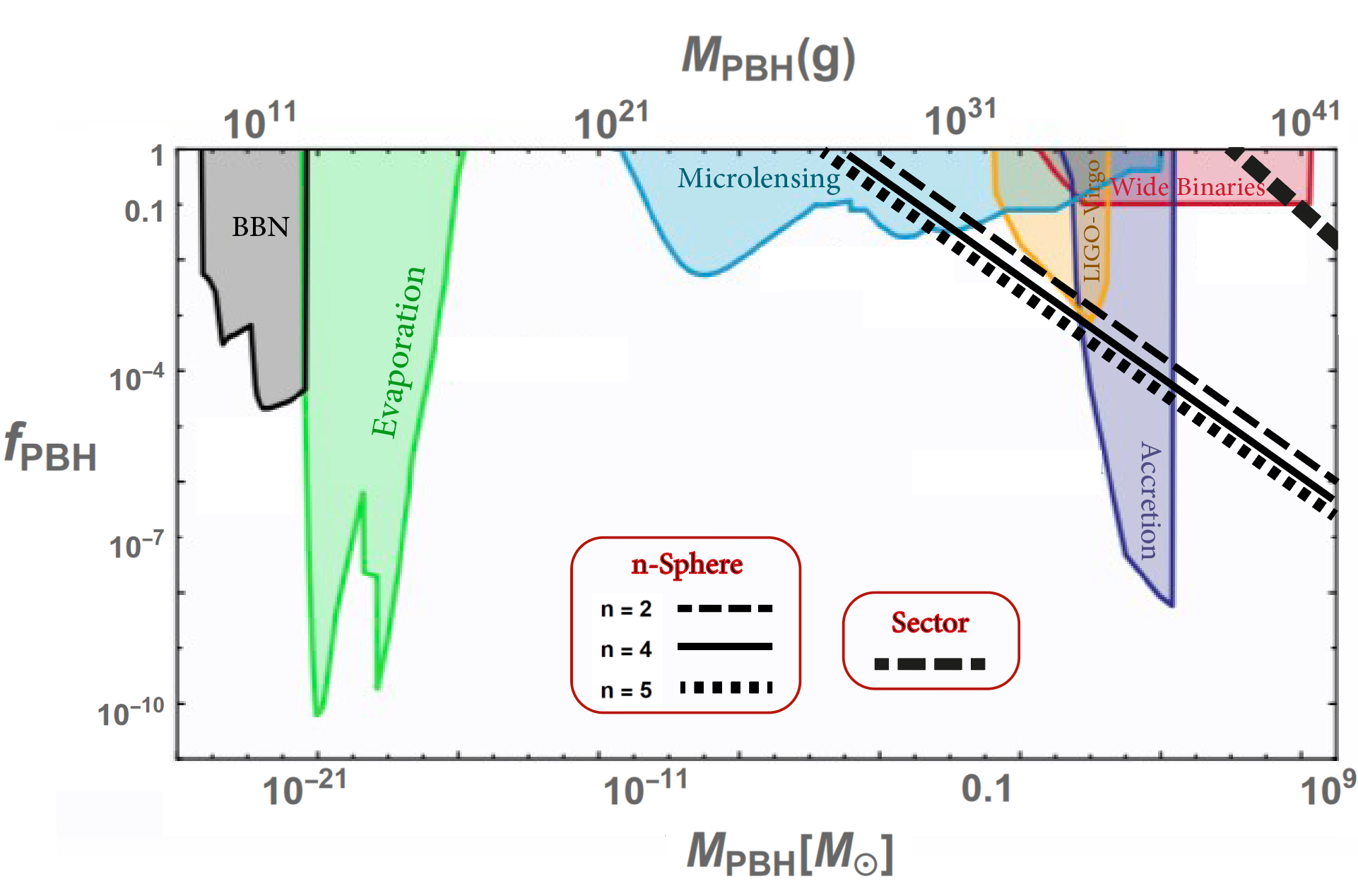}
    \caption{The behaviour of $f_{PBH}$ as a function of the mass of the formed PBHs (in gr as well as  in unit of solar mass) for both spherical boundaries with various values of the field space dimensions $n$ and the sector boundary .}
    \label{f_PBH}
\end{figure}

As mentioned before, another important parameter in studying the PBHs of a model is the fraction of PBHs against the total dark matter density, $f_{PBH}$. Fig. \ref{f_PBH} shows the behaviour of $f_{PBH}$ versus $M_{PBH}[\textup{M}_\odot]$ for various values of $n$ for the case of spherical boundaries as well as for the case of the sector.  
The observational constraints presented in this figure are obtained with the help of the PBHbounds package \cite{PBH:BoundsGit, Passaglia:2021jla}. As can be seen, these constraints are applied on the mass of PBHs from $10^9 $ gr to $10^{42}$ gr which is shown in the unit of ${M}_\odot$ as well.   We emphasize that constraints used in Fig. \ref{f_PBH} are not necessarily the strongest constraints across their respective mass ranges. Moreover, in Fig. \ref{f_PBH} we study $f_{PBH}$ for spherical boundaries with  $n=2$, $n=4$ and $n=5$ while the  result of mass fraction for $n=3$ is not presented since no values of the parameters can give a valid range of $\beta$. The general prediction of our setup is that the resultant PBHs are typically heavier than  the solar mass with $f_{PBH} $ at the order $10^{-2}- 10^{-5}$.

\section{Summary and Discussions}
\label{conclusion}

In this paper we have studied multiple fields inflation in the diffusion dominated regime using the stochastic $\delta N$ formalism. In order to terminate inflation,  boundaries in field space with various boundary conditions have been imposed. Although this setup in which the fields are under pure Brownian motion is idealized  it can mimic some limits of physical interests in inflationary model building. Our main motivation for this purpose was to consider the multiple fields USR setup in its final stages. During the USR phase the classical velocity of the fields fall off exponentially so after a few e-folds one may safely neglect the classical drift terms  and the main source of the evolution of the fields are given by quantum diffusion terms. This is the idealized limit which we studied in this paper. As for the boundaries, we have allowed for both reflective and absorbing boundary conditions. The former is typically imposed in the UV region so one demands that the fields do not explore the far UV region, i.e. the quantum gravity limit. On the other hand, the absorbing boundaries may be interpreted as the surfaces of end of inflation (or end of USR).
We comment that although our main motivation was to study the idealized version of USR setup,  our study of quantum diffusions in multi dimensional field space can be very useful in the context of eternal inflation in higher dimensional landscape. 
This can happen when some regions of the higher dimensional landscape are flat enough so the  dynamics of fields is determined by the quantum diffusion kicks.

The geometries we studied include both the  symmetric and asymmetric boundaries. In the symmetric configurations we considered two concentric 
$n-1$ dimensional spheres as the boundaries in the $n$ dimensional field space. In order to terminate inflation, we require at least one boundary to be absorbing. We have calculated the 
mean number of e-folds $\langle \cN \rangle$ for the duration of USR. If both boundaries are absorbing, we also have calculated the first hitting probabilities $p_\pm$ to terminate inflation 
which agrees with the results in  \cite{Assadullahi:2016gkk} in the limit where they overlap.  On the other hand, when one boundary is absorbing and the other one is reflecting, there is no notion of first crossing probabilities but instead, we can calculate the curvature perturbation power spectrum $\calP_\calR$. One may ask how the power spectrum changes as a function of the  dimension of field space. 
As we have shown in Fig. \ref{power_Circle}, this depends on the relative positions of the two boundaries. Suppose we start with a fixed value of the initial radial position $r$ in field space for all values of $n$.  if the interior boundary $r_-$ is absorbing and the outer boundary $r_+$ is reflective, then by increasing the dimension of the field space, the power spectrum increases as well. This trend is reversed when the positions of the absorbing and reflective boundaries are switched. Another general conclusion of our study was that by increasing the dimension of the field space, there will be more volume for the fields to wander around. Consequently, with fixed values of the initial radial position, it becomes  more likely that the field hits the large exterior boundary than the small interior boundary.

We have extended our studies to the case where the boundaries are asymmetric as well.  As the analysis are more complicated we set $n=2$ and studied the rectangle and the sector boundaries as some case studies.  As in the case of symmetric boundaries, we have calculated $\langle \cN \rangle, p_\pm$ and $\calP_\calR$.
One additional complexity associated with asymmetric boundaries  was that since both the radial and angular positions  are independent variables, then there is ambiguities in the definition of power spectrum. In other words, the power spectrum depends on the trajectory in field space which connects the initial position to the final position on the absorbing boundary.  As simple examples to calculate $\calP_\calR$,  for the case of rectangle boundaries, we considered the trajectories to be perpendicular to the absorbing boundary while for the case of sector we considered a pure angular trajectory with a fixed radial position along the path.

As a cosmological application, we have studied the PBHs formation within our setup for  various cases.  As in single field USR setup, one expects that PBHs to form in this setup which may comprise all or part of the dark matter energy density. We have shown that there are regions in our parameter space where PBHs with mass fraction  $10^{-12}<\beta(M)<10^{-11}$ and various mass ranges can be generated. However, our model typically predicts PBHs with $M > \textup{M}_\odot$ which can only furnish a relatively small fraction of the dark matter, say $f_{PBH} \sim 10^{-2} - 10^{-5} $. 

There are a number of directions which the current studies can be extended. One question is to study asymmetric boundaries with more realistic configurations. This includes the ellipsoid and hyperbola boundaries in field space. Another question is 
to study models with a curved field space. In our current analysis the field space is flat while the boundaries are curved (for $n>2$). Finally, an important and physically more relevant question is to consider the case where the drift also plays important roles. This corresponds to the early stage of multiple fields  
USR inflation where the  classical drifts have not fallen exponentially. 
We would like to come back to this question and its cosmological implications  in future. 

\vspace{1 cm}

{\bf Acknowledgments:} We  would like to thank S. Hooshangi and  A. Talebian for helpful discussions about PBHs formations.
H. F. would like to thank ``Saramadan'' federation of Iran 
for partial support. 

\vspace{0.5cm}

\appendix
\section{PDF from the Fokker Planck Equation }\label{PDF-Sector}

In this Appendix we obtain the PDF for the case of sector boundary from the Fokker-Planck equation. 

In general one can  start by the Fokker-Planck equation which, in the diffusion dominated regime, is written as follows:
\begin{equation}
    \frac{1}{2}\nabla'^2P(x',x,N-N_i)=\frac{\partial P(x',x,N-N_i)}{\partial N},
\end{equation}
where $N_i$ and $x$ are  the initial time and position of the field in 
$n$-dimensional space respectively. For simplicity we set $N_i=0$. The appropriate boundary conditions for  $P(x',x,N)$ are as follow:

\begin{equation}\label{delta}
    P(x',x,0)=\delta(x'-x) , 
\end{equation}
and 
\begin{equation}
    \nabla' P(x',x,N).\hat{n}\big|_{x'=R_i}=0 \, , \quad \quad 
    P(A_i,x,N)=0,
\end{equation}
where $A_i$ and $R_i$ are the $i-$th absorbing and reflective boundaries respectively. Moreover, $\hat{n}$ is the orthonormal vector on the $R_i$. 

Now  suppose the sector with radius $R$ has mixed boundary conditions. In other words the radial edge at $\theta'=0$ and the bow with $r=R$ are reflective while the radial edge at $\theta'=\alpha$ is absorbing. Then by the  method of separation of variables one can propose the following solution for the Fokker-Planck equation
\begin{equation}
\begin{split}
  P(r',\theta',\theta,r,N)=  \sum_{m=0,k_m}^\infty &\Bigg[A_{mk} J_l(k_m r')+B_{mk}Y_l(k_m r') \Bigg]\cos(l \theta')\exp(-\frac{r_{mk}^2N}{2}) \, ,
  \end{split}
\end{equation}
where $l = \frac{(2m+1)\pi}{2\alpha}$. In the limit that $r'=0$ , $P(r'=0,\theta',\theta,r,N)=0$ and so $B_m=0$.  One can now determine $k$ and $A_m$ from the other boundary conditions. Since on the bow we should have $\frac{\partial P(r',\theta',\theta,r,N)}{\partial r'}\big|_{r'=R}=0$ then  $k_m$ is proportional to the $k$-th root of the derivative of  $J_{l}$  which we denote  by $r_{m k}$. Employing   \eqref{delta} in the polar coordinates we obtain  
\begin{equation}
\begin{split}
  P(r',\theta',\theta,r,N)=  \sum_{m=0,k=1}^\infty &\Big[A_{mk} J_{l}\Big(\frac{r_{mk}r'}{R}\Big)\Big]\cos(l\theta')\exp(-\frac{r_{mk}^2N}{2}) \, ,
  \end{split}
\end{equation}
where $A_{mk}$ is given by
\begin{equation}
    A_{mk}= -\frac{4}{\alpha R^2} \frac{J_{l}(\frac{r_{mk}r}{R})\cos(l\theta)}{J''_{l}(r_{mk})J_{l}(r_{mk})} \, ,
\end{equation}
in which we have used the following orthogonality relation for the Bessel functions satisfying the Neumann boundary conditions in the sector
\begin{equation}
   \int_0^{R}rJ_{l}\Big(\frac{r_{mk}r}{R}\Big)J_{l}\Big(\frac{r_{mk'}r}{R}\Big)dr=-\delta_{kk'}\frac{R^2}{2}J''_{l}(r_{mk})J_{l}(r_{mk}).
\end{equation}
Having the solution to the Fokker-Planck equation one can obtain the first crossing time PDF using 
\begin{equation}\label{PDFfokker}
    P(\mathcal{N})=-\frac{\partial}{\partial N}\int_A P(x',x,N)dA,
\end{equation}
where we have integrated over the surface of the two dimensional bulk. Using the Fokker-Planck equation one may write the above relation as
\begin{equation}
     P(\mathcal{N})=-\frac{1}{2}\int_A \nabla'^2P(x',x,N)dA \, ,
\end{equation}
By using the Stokes theorem one may write
\begin{equation}
   -\frac{1}{2}\int_A \nabla'^2P(x',x,N)dA=-\frac{1}{2}\oint\big( -\frac{\partial P}{\partial y}dy+\frac{\partial P}{\partial x}dx\big) \, .
\end{equation}
As the sector has three edges then the above expression reduces to the following equation in the polar coordinates:
\begin{equation}
\begin{split}
    -\frac{1}{2}\oint\big( -\frac{\partial P}{\partial y}dy+\frac{\partial P}{\partial x}dx\big)=-\frac{1}{2}&\Big[-\int_{C_1}\frac{1}{r}\frac{\partial P(r',\theta',r,\theta)}{\partial \theta'}\big|_{\theta'=0}dr'-\\ &\int_{C_2}\frac{1}{r}\frac{\partial P(r',\theta',r,\theta)}{\partial \theta'}\big|_{\theta'=\alpha}dr'+R\int_{C_3}\frac{\partial P(r',\theta',r,\theta)}{\partial r'}\big |_{r'=R}d\theta'\Big].
    \end{split}
\end{equation}
Each component in the right hand side of  the above equation defines a current probability on the $C_i$ boundary by which we can define the PDF of crossing the $i$-th boundary. So we write:

\begin{equation}
    J_{C_i}=-\frac{1}{2}\int_{C_i}\big( -\frac{\partial P}{\partial y}dy+\frac{\partial P}{\partial x}dx) \, ,
\end{equation}
where the integral is taken counter-clockwise. The current probability may be easily generalized to higher dimensions using generalized Stokes theorem. The PDF is then given by 
\begin{equation}
    P(\mathcal{N})=\sum_{i=1}^N J_{C_i}(N) \, .
\end{equation} 

As in the case of sector with mixed boundary conditions, $C_1$ and $C_3$ are reflective then the current probability vanishes on them and we may write
\begin{equation}
    P(\mathcal{N})=J_{C_2}
\end{equation}
which yields to   Eq.\eqref{PDFsector}.


\vspace{0.5cm}

\begin{thebibliography}{}


\bibitem{Weinberg:2008zzc}
S.~Weinberg,
``Cosmology,'' Oxford University press, 2008.

\bibitem{Baumann:2009ds}
D.~Baumann,
``Inflation,''
[arXiv:0907.5424 [hep-th]].

\bibitem{Planck:2018vyg}
N.~Aghanim \textit{et al.} [Planck],
Astron. Astrophys. \textbf{641}, A6 (2020)
[erratum: Astron. Astrophys. \textbf{652}, C4 (2021)], 


\bibitem{Planck:2018jri}
Y.~Akrami \textit{et al.} [Planck],
Astron. Astrophys. \textbf{641}, A10 (2020).

\bibitem{Wands:2007bd}
D.~Wands,
``Multiple field inflation,''
Lect. Notes Phys. \textbf{738}, 275-304 (2008). 

\bibitem{Dimopoulos:2005ac}
S.~Dimopoulos, S.~Kachru, J.~McGreevy and J.~G.~Wacker,
JCAP \textbf{08}, 003 (2008).

\bibitem{Ashoorioon:2009wa}
A.~Ashoorioon, H.~Firouzjahi and M.~M.~Sheikh-Jabbari,
JCAP \textbf{06}, 018 (2009). 

\bibitem{Kodama:1984ziu}
H.~Kodama and M.~Sasaki,
Prog. Theor. Phys. Suppl. \textbf{78}, 1-166 (1984).

\bibitem{Weinberg:2005vy}
S.~Weinberg,
Phys. Rev. D \textbf{72}, 043514 (2005).



\bibitem{Vilenkin:1983xp} 
A.~Vilenkin, 
Nucl.\ Phys.\ B {\bf 226}, 527 (1983).


\bibitem{Starobinsky:1986fx} 
  A.~A.~Starobinsky,
  Lect.\ Notes Phys.\  {\bf 246}, 107 (1986).

  \bibitem{Starobinsky:1994bd}
A.~A. Starobinsky and J.~Yokoyama, 
  { Phys.Rev.} {\bf D50} (1994)
  6357--6368.


\bibitem{Rey:1986zk}
S.~J.~Rey,
Nucl. Phys. B \textbf{284}, 706-728 (1987).

\bibitem{Nakao:1988yi}
K.-i. Nakao, Y.~Nambu, and M.~Sasaki, 
{ Prog.Theor.Phys.} {\bf 80} (1988) 1041.
  
\bibitem{Sasaki:1987gy} 
  M.~Sasaki, Y.~Nambu and K.~i.~Nakao,
  Nucl.\ Phys.\ B {\bf 308}, 868 (1988).
  
  
\bibitem{Kandrup:1988sc}
H.~E. Kandrup, 
  { Phys.Rev.} {\bf D39} (1989) 2245.
\bibitem{Nambu:1987ef}
Y.~Nambu and M.~Sasaki, 
  { Phys.Lett.} {\bf B205} (1988) 441.

\bibitem{Nambu:1988je}
Y.~Nambu and M.~Sasaki, 
  { Phys.Lett.} {\bf B219} (1989) 240.


\bibitem{Nambu:1989uf}
Y.~Nambu, 
{ Prog.Theor.Phys.} {\bf 81} (1989) 1037.

\bibitem{Mollerach:1990zf}
S.~Mollerach, S.~Matarrese, A.~Ortolan, and F.~Lucchin, 
{ Phys.Rev.} {\bf D44} (1991) 1670--1679.

\bibitem{Linde:1993xx}
A.~D. Linde, D.~A. Linde, and A.~Mezhlumian, 
  { Phys.Rev.} {\bf D49} (1994)
  1783--1826. 

  
\bibitem{Kunze:2006tu} 
  K.~E.~Kunze,
  JCAP {\bf 0607}, 014 (2006). 

\bibitem{Prokopec:2007ak} 
  T.~Prokopec, N.~C.~Tsamis and R.~P.~Woodard,
  Annals Phys.\  {\bf 323}, 1324 (2008). 
  
\bibitem{Prokopec:2008gw} 
  T.~Prokopec, N.~C.~Tsamis and R.~P.~Woodard,
  Phys.\ Rev.\ D {\bf 78}, 043523 (2008). 
 
\bibitem{Tsamis:2005hd} 
  N.~C.~Tsamis and R.~P.~Woodard,
  Nucl.\ Phys.\ B {\bf 724}, 295 (2005). 

\bibitem{Enqvist:2008kt} 
  K.~Enqvist, S.~Nurmi, D.~Podolsky and G.~I.~Rigopoulos,
  JCAP {\bf 0804}, 025 (2008). 
  

\bibitem{Finelli:2008zg}
F.~Finelli, G.~Marozzi, A.~Starobinsky, G.~Vacca, and G.~Venturi, 
  { Phys.Rev.} {\bf D79} (2009) 044007.

\bibitem{Finelli:2010sh}
F.~Finelli, G.~Marozzi, A.~Starobinsky, G.~Vacca, and G.~Venturi, 
  { Phys.Rev.} {\bf D82} (2010) 064020.

\bibitem{Garbrecht:2013coa}
B.~Garbrecht, G.~Rigopoulos, and Y.~Zhu, 
{ Phys.Rev.} {\bf D89} (2014) 063506. 

\bibitem{Garbrecht:2014dca}
B.~Garbrecht, F.~Gautier, G.~Rigopoulos, and Y.~Zhu, 
{ Phys. Rev.} {\bf D91} (2015), no.~6 063520. 

\bibitem{Burgess:2014eoa} 
  C.~P.~Burgess, R.~Holman, G.~Tasinato and M.~Williams,
  JHEP {\bf 1503}, 090 (2015). 

\bibitem{Burgess:2015ajz} 
  C.~P.~Burgess, R.~Holman and G.~Tasinato,
  JHEP {\bf 1601}, 153 (2016). 

\bibitem{Boyanovsky:2015tba} 
  D.~Boyanovsky,
  Phys.\ Rev.\ D {\bf 92}, no. 2, 023527 (2015).
  
\bibitem{Boyanovsky:2015jen} 
  D.~Boyanovsky,
  Phys.\ Rev.\ D {\bf 93}, 043501 (2016). 
  

\bibitem{Pinol:2020cdp}
L.~Pinol, S.~Renaux-Petel and Y.~Tada,
JCAP \textbf{04}, 048 (2021). 

\bibitem{Cruces:2018cvq}
D.~Cruces, C.~Germani and T.~Prokopec,
JCAP \textbf{03}, 048 (2019). 

\bibitem{Cruces:2021iwq}
D.~Cruces and C.~Germani,
Phys. Rev. D \textbf{105}, no.2, 023533 (2022). 


\bibitem{Noorbala:2019kdd}
M.~Noorbala and H.~Firouzjahi,
Phys. Rev. D \textbf{100}, no.8, 083510 (2019). 


\bibitem{Ahmadi:2022lsm}
N.~Ahmadi, M.~Noorbala, N.~Feyzabadi, F.~Eghbalpoor and Z.~Ahmadi,
JCAP \textbf{08}, 078 (2022). 


\bibitem{Pattison:2019hef}
C.~Pattison, V.~Vennin, H.~Assadullahi and D.~Wands,
JCAP \textbf{07}, 031 (2019). 

\bibitem{Pattison:2021oen}
C.~Pattison, V.~Vennin, D.~Wands and H.~Assadullahi,
JCAP \textbf{04}, 080 (2021).

\bibitem{Firouzjahi:2018vet}
H.~Firouzjahi, A.~Nassiri-Rad and M.~Noorbala,
JCAP \textbf{01}, 040 (2019).


\bibitem{Firouzjahi:2020jrj}
H.~Firouzjahi, A.~Nassiri-Rad and M.~Noorbala,
Phys. Rev. D \textbf{102} (2020) no.12, 123504.


\bibitem{Mishra:2023lhe}
S.~S.~Mishra, E.~J.~Copeland and A.~M.~Green,
[arXiv:2303.17375 [astro-ph.CO]].

\bibitem{Fujita:2017lfu} 
  T.~Fujita and I.~Obata,
  JCAP {\bf 1801}, no. 01, 049 (2018). 
  
\bibitem{Fujita:2022fit}
T.~Fujita, K.~Mukaida and Y.~Tada,
[arXiv:2206.12218 [astro-ph.CO]].

\bibitem{Fujita:2022fwc}
T.~Fujita, J.~Kume, K.~Mukaida and Y.~Tada,
[arXiv:2204.01180 [hep-ph]].

\bibitem{Talebian:2019opf}
A.~Talebian, A.~Nassiri-Rad and H.~Firouzjahi,
Phys. Rev. D \textbf{101}, no.2, 023524 (2020).

\bibitem{Talebian:2020drj}
A.~Talebian, A.~Nassiri-Rad and H.~Firouzjahi,
Phys. Rev. D \textbf{102}, no.10, 103508 (2020). 

\bibitem{Talebian:2021dfq}
A.~Talebian, A.~Nassiri-Rad and H.~Firouzjahi,
Phys. Rev. D \textbf{105}, no.2, 023528 (2022). 

\bibitem{Talebian:2022jkb}
A.~Talebian, A.~Nassiri-Rad and H.~Firouzjahi,
Phys. Rev. D \textbf{105}, no.10, 103516 (2022).


\bibitem{Fujita:2013cna}
  T.~Fujita, M.~Kawasaki, Y.~Tada and T.~Takesako,
  JCAP {\bf 1312}, 036 (2013).

\bibitem{Fujita:2014tja}
  T.~Fujita, M.~Kawasaki and Y.~Tada,
  JCAP {\bf 1410}, no. 10, 030 (2014).

\bibitem{Vennin:2015hra}
  V.~Vennin and A.~A.~Starobinsky,
  Eur.\ Phys.\ J.\ C {\bf 75}, 413 (2015).

\bibitem{Vennin:2016wnk}
  V.~Vennin, H.~Assadullahi, H.~Firouzjahi, M.~Noorbala and D.~Wands,
  Phys.\ Rev.\ Lett.\  {\bf 118}, no. 3, 031301 (2017).

\bibitem{Assadullahi:2016gkk}
  H.~Assadullahi, H.~Firouzjahi, M.~Noorbala, V.~Vennin and D.~Wands,
  JCAP {\bf 1606}, no. 06, 043 (2016).


\bibitem{Grain:2017dqa}
  J.~Grain and V.~Vennin,
  JCAP {\bf 1705}, no. 05, 045 (2017).

\bibitem{Noorbala:2018zlv}
  M.~Noorbala, V.~Vennin, H.~Assadullahi, H.~Firouzjahi and D.~Wands,
  JCAP {\bf 1809}, no. 09, 032 (2018).

\bibitem{Jackson:2022unc}
J.~H.~P.~Jackson, H.~Assadullahi, K.~Koyama, V.~Vennin and D.~Wands,
[arXiv:2206.11234 [astro-ph.CO]].





\bibitem{Sasaki:1995aw}
  M.~Sasaki and E.~D.~Stewart,
  Prog.\ Theor.\ Phys.\  {\bf 95}, 71 (1996).

\bibitem{Sasaki:1998ug}
  M.~Sasaki and T.~Tanaka,
  Prog.\ Theor.\ Phys.\  {\bf 99}, 763 (1998).

\bibitem{Lyth:2004gb}
  D.~H.~Lyth, K.~A.~Malik and M.~Sasaki,
  JCAP {\bf 0505}, 004 (2005).

\bibitem{Wands:2000dp}
  D.~Wands, K.~A.~Malik, D.~H.~Lyth and A.~R.~Liddle,
  Phys.\ Rev.\ D {\bf 62}, 043527 (2000).

\bibitem{Lyth:2005fi}
  D.~H.~Lyth and Y.~Rodriguez,
  Phys.\ Rev.\ Lett.\  {\bf 95}, 121302 (2005).

\bibitem{Abolhasani:2019cqw}
  A.~A.~Abolhasani, H.~Firouzjahi, A.~Naruko and M.~Sasaki,
  doi:10.1142/10953

\bibitem{Abolhasani:2018gyz}
  A.~A.~Abolhasani and M.~Sasaki,
  JCAP {\bf 1808}, no. 08, 025 (2018).




\bibitem{Hawking:1971mnras}
S. Hawking, Gravitationally collapsed objects of very low mass, Mon. Not. Roy. Astron.
Soc. 152 (1971) 75.

\bibitem{Carr:1974 mnras}
B. J. Carr and S. W. Hawking, Black holes in the early Universe, Mon. Not. Roy. Astron. Soc. 168 (1974) 399-415.

\bibitem{CarrPBH:1975 Apj}
B. J. Carr, The Primordial black hole mass spectrum, Astrophys. J. 201 (1975) 1-19.

\bibitem{Sasaki:2018dmp}
M.~Sasaki, T.~Suyama, T.~Tanaka and S.~Yokoyama,
Class. Quant. Grav. \textbf{35}, no.6, 063001 (2018). 

\bibitem{Carr:2020gox}
B.~Carr, K.~Kohri, Y.~Sendouda and J.~Yokoyama,
Rept. Prog. Phys. \textbf{84}, no.11, 116902 (2021). 

\bibitem{Green:2020jor}
A.~M.~Green and B.~J.~Kavanagh,
J. Phys. G \textbf{48}, no.4, 043001 (2021).


\bibitem{Byrnes:2021jka}
C.~T.~Byrnes and P.~S.~Cole,
[arXiv:2112.05716 [astro-ph.CO]].


\bibitem{Kinney:2005vj} 
  W.~H.~Kinney,
  Phys.\ Rev.\ D {\bf 72}, 023515 (2005).
  
  
\bibitem{Namjoo:2012aa} 
  M.~H.~Namjoo, H.~Firouzjahi and M.~Sasaki,
  EPL {\bf 101}, no. 3, 39001 (2013). 

\bibitem{Chen:2013aj}
X.~Chen, H.~Firouzjahi, M.~H.~Namjoo and M.~Sasaki,
EPL \textbf{102}, no.5, 59001 (2013). 
  
\bibitem{Martin:2012pe} 
  J.~Martin, H.~Motohashi and T.~Suyama,
  Phys.\ Rev.\ D {\bf 87}, no. 2, 023514 (2013).

\bibitem{Akhshik:2015rwa}
M.~Akhshik, H.~Firouzjahi and S.~Jazayeri,
JCAP \textbf{12}, 027 (2015). 



\bibitem{Nassiri-Rad:2022azj}
A.~Nassiri-Rad, K.~Asadi and H.~Firouzjahi,
Phys. Rev. D \textbf{106}, no.12, 123528 (2022). 


\bibitem{Jackson}
Jackson, John David, 1925-2016. Classical Electrodynamics. New York :Wiley, 1999.

\bibitem{LIGOScientific:2016aoc}
B.~P.~Abbott \textit{et al.} [LIGO Scientific and Virgo],
Phys. Rev. Lett. \textbf{116}, no.6, 061102 (2016). 

\bibitem{LIGOScientific:2016sjg}
B.~P.~Abbott \textit{et al.} [LIGO Scientific and Virgo],
Phys. Rev. Lett. \textbf{116}, no.24, 241103 (2016)

\bibitem{Carr:2016drx}
B.~Carr, F.~Kuhnel and M.~Sandstad,
Phys. Rev. D \textbf{94}, no.8, 083504 (2016). 



\bibitem{Ivanov:1994pa}
P.~Ivanov, P.~Naselsky and I.~Novikov,
Phys. Rev. D \textbf{50}, 7173-7178 (1994) .


\bibitem{Garcia-Bellido:2017mdw}
J.~Garcia-Bellido and E.~Ruiz Morales,
Phys. Dark Univ. \textbf{18}, 47-54 (2017).

\bibitem{Biagetti:2018pjj}
M.~Biagetti, G.~Franciolini, A.~Kehagias and A.~Riotto,
JCAP \textbf{07}, 032 (2018). 

\bibitem{Franciolini:2018vbk}
G.~Franciolini, A.~Kehagias, S.~Matarrese and A.~Riotto,
JCAP \textbf{03}, 016 (2018). 

\bibitem{Motohashi:2017kbs}
H.~Motohashi and W.~Hu,
Phys. Rev. D \textbf{96}, no.6, 063503 (2017). 

\bibitem{Germani:2017bcs}
C.~Germani and T.~Prokopec,
Phys. Dark Univ. \textbf{18}, 6-10 (2017). 

\bibitem{Ragavendra:2020sop}
H.~V.~Ragavendra, P.~Saha, L.~Sriramkumar and J.~Silk,
Phys. Rev. D \textbf{103}, no.8, 083510 (2021). 


\bibitem{Ozsoy:2021pws}
O.~\"Ozsoy and G.~Tasinato,
Phys. Rev. D \textbf{105}, no.2, 023524 (2022). 

\bibitem{Hooshangi:2022lao}
S.~Hooshangi, A.~Talebian, M.~H.~Namjoo and H.~Firouzjahi,
Phys. Rev. D \textbf{105}, no.8, 8 (2022). 

\bibitem{Cai:2022erk}
Y.~F.~Cai, X.~H.~Ma, M.~Sasaki, D.~G.~Wang and Z.~Zhou,
JCAP \textbf{12}, 034 (2022). 

\bibitem{Cai:2021zsp}
Y.~F.~Cai, X.~H.~Ma, M.~Sasaki, D.~G.~Wang and Z.~Zhou,
Phys. Lett. B \textbf{834}, 137461 (2022). 


\bibitem{Pi:2022ysn}
S.~Pi and M.~Sasaki,
[arXiv:2211.13932 [astro-ph.CO]].


\bibitem{Kristiano:2022maq}
J.~Kristiano and J.~Yokoyama,
[arXiv:2211.03395 [hep-th]].

\bibitem{Kristiano:2023scm}
J.~Kristiano and J.~Yokoyama,
[arXiv:2303.00341 [hep-th]].

\bibitem{Cheng:2021lif}
S.~L.~Cheng, D.~S.~Lee and K.~W.~Ng,
Phys. Lett. B \textbf{827}, 136956 (2022). 


\bibitem{Riotto:2023gpm}
A.~Riotto,
[arXiv:2303.01727 [astro-ph.CO]].

\bibitem{Riotto:2023hoz}
A.~Riotto,
[arXiv:2301.00599 [astro-ph.CO]].

\bibitem{Choudhury:2023vuj}
S.~Choudhury, M.~R.~Gangopadhyay and M.~Sami,
[arXiv:2301.10000 [astro-ph.CO]].


\bibitem{Choudhury:2023jlt}
S.~Choudhury, S.~Panda and M.~Sami,
[arXiv:2302.05655 [astro-ph.CO]].


\bibitem{Choudhury:2023rks}
S.~Choudhury, S.~Panda and M.~Sami,
[arXiv:2303.06066 [astro-ph.CO]].

\bibitem{Firouzjahi:2023aum}
H.~Firouzjahi,
[arXiv:2303.12025 [astro-ph.CO]].

\bibitem{Ezquiaga:2019ftu}
J.~M.~Ezquiaga, J.~Garc\'\i{}a-Bellido and V.~Vennin,
JCAP \textbf{03}, 029 (2020). 


\bibitem{Carr:2020}
B.~Carr, K.~Kohri, Y.~Sendouda and J.~Yokoyama,
Rept. Prog. Phys. \textbf{84}, no.11, 116902 (2021). 





\bibitem{Carr:2009}
B.~J.~Carr, K.~Kohri, Y.~Sendouda and J.~Yokoyama,
Phys. Rev. D \textbf{81}, 104019 (2010).




\bibitem{Carr:2017}
B.~Carr, M.~Raidal, T.~Tenkanen, V.~Vaskonen and H.~Veerm\"ae,
Phys. Rev. D \textbf{96}, no.2, 023514 (2017)




\bibitem{PBH:BoundsGit}
B. J. Kavanagh et al., "PBHBounds", GitHub repository.

\bibitem{Passaglia:2021jla}
S.~Passaglia and M.~Sasaki,
Phys. Rev. D \textbf{105}, no.10, 103530 (2022)




\end{thebibliography}
\end{document}